\documentclass[twocolumn]{aastex62}
\usepackage{natbib}
\graphicspath{{./}{figures/}}
\usepackage{color,subfigure,lineno,booktabs} 
\definecolor{firebrick}{rgb}{0.7, 0.13, 0.13}
\newcommand{\fire}[1]{\textcolor{firebrick}{\textbf{#1}}}

\newcommand{\hbeta}{H{$\beta$}}
\newcommand{\halpha}{H{$\alpha$}}

\def\CIV{C\,{\sc iv}}
\def\CIII{C\,{\sc iii]}}
\def\HeII{He\,{\sc ii}}
\def\CIV{C\,{\sc iv}}
\def\MgII{Mg\,{\sc ii}}

\def\OIII{[O\,{\sc iii}]}
\def\NeV{[Ne\,{\sc v}]}
\def\HeII{He\,{\sc ii}}

\def\OII{[O\,{\sc ii}]}
\def\NeIII{[Ne\,{\sc iii}]}

\def\proj{\texttt{NEXUS}}

\shorttitle{NEXUS: Overview} 
\shortauthors{Shen et~al.}
\begin{document}

\title{NEXUS: the North ecliptic pole EXtragalactic Unified Survey}

%\correspondingauthor{}
%\email{}

\author[0000-0003-1659-7035]{Yue Shen}
\affiliation{Department of Astronomy, University of Illinois Urbana-Champaign, Urbana, IL 61801, USA}
\affiliation{National Center for Supercomputing Applications, University of Illinois Urbana-Champaign, Urbana, IL 61801, USA}

\author[0000-0001-5105-2837]{Ming-Yang Zhuang}
\affiliation{Department of Astronomy, University of Illinois Urbana-Champaign, Urbana, IL 61801, USA}

\author[0000-0002-1605-915X]{Junyao Li}
\affiliation{Department of Astronomy, University of Illinois Urbana-Champaign, Urbana, IL 61801, USA}

\author[0000-0002-6523-9536]{Adam J.~Burgasser}
\affiliation{Department of Astronomy \& Astrophysics, UC San Diego, La Jolla, CA, USA}

\author[0000-0003-3310-0131]{Xiaohui Fan} 
\affiliation{Steward Observatory, University of Arizona, Tucson, AZ 85750, USA}

\author[0000-0002-5612-3427]{Jenny E.~Greene}
\affiliation{Department of Astrophysical Sciences, Princeton University, Princeton, NJ 08544, USA}

\author[0000-0001-6022-0484]{Gautham Narayan}
\affiliation{Department of Astronomy, University of Illinois Urbana-Champaign, Urbana, IL 61801, USA}
\affiliation{Center for AstroPhysical Surveys, National Center for Supercomputing Applications, University of Illinois Urbana-Champaign, Urbana, IL 61801, USA}

\author[0000-0003-3509-4855]{Alice E.~Shapley}
\affiliation{Department of Physics \& Astronomy, University of California, Los Angeles, 430 Portola Plaza, Los Angeles, CA 90095, USA}

\author[0000-0002-4622-6617]{Fengwu Sun}
\affiliation{Center for Astrophysics $\vert$ Harvard \& Smithsonian, 60 Garden Street, Cambridge, MA 02138, USA}

\author[0000-0002-7633-431X]{Feige Wang}
\affiliation{Steward Observatory, University of Arizona, Tucson, AZ 85750, USA}

\author[0000-0002-6893-3742]{Qian Yang}
\affiliation{Center for Astrophysics $\vert$ Harvard \& Smithsonian, 60 Garden Street, Cambridge, MA 02138, USA}

\begin{abstract}
NEXUS is a JWST Multi-Cycle (Cycles 3--5; 368 primary hrs) GO Treasury imaging and spectroscopic survey around the North Ecliptic Pole. It contains two overlapping tiers. The Wide tier ($\sim 400~{\rm arcmin}^2$) performs NIRCam/WFSS 2.4--5\,\micron\ grism spectroscopy with three epochs over 3 years (final continuum ${\rm S/N/pixel>3}$ at F444W$<22.2$). The Deep tier ($\sim 50~{\rm arcmin}^2$) performs high-multiplexing NIRSpec 0.6--5.3\,\micron\ MOS/PRISM spectroscopy for $\sim 10,000$ targets, over 18 epochs with a 2-month cadence (epoch/final continuum ${\rm S/N/pixel>3}$ at F200W$\lesssim 27/29$). All epochs have simultaneous multi-band NIRCam and MIRI imaging ($5\sigma$ final depths of $\sim 28-29$ in NIRCam and $\sim 25$ in MIRI). The field is within the continuous viewing zone of JWST, and is fully covered by the Euclid Ultra-Deep Field, with 0.9--2\,\micron\ deep Euclid spectroscopy and cadenced photometry to maximize synergy across wavelengths and science areas. NEXUS has three science pillars. First, with its massive and nearly complete (flux-limited) spectroscopic samples and deep photometry, it will perform efficient classification and physical characterization of galaxies and AGNs from $z\sim 1$ to Cosmic Dawn. With the large contiguous area coverage, it will measure the spatial clustering and demography of the first galaxies and SMBHs at $z>6$. Second, multi-epoch observations enable systematic time-domain investigations, focusing on $z>3$ transients and low-mass AGN reverberation mapping. Third, the comprehensive data set will enable knowledge transfer to other legacy fields, create data challenges, and initiate benchmark work for future space missions. With rapid public releases of processed data and an open invitation for collaboration, NEXUS aims for broad and swift community engagement, to become a powerhouse to drive transformative advancements in multiple key science areas of astronomy.
\end{abstract}

\section{Introduction}\label{sec:introduction}

The first wave of JWST imaging and spectroscopy has started revolutionizing many fields in astronomy. These deep, IR-wavelength data have led to the discovery of an abundance of high-redshift (e.g., $z\gtrsim 6-14$) galaxies and AGNs, pushing into the low-mass regime occupied by the fast-growing galaxy population at Cosmic Dawn ($z\gtrsim 6$). JWST survey programs, e.g., JADES \citep{JADES}, UNCOVER \citep{UNCOVER}, COSMOS-Web \citep{COSMOS-Web}, FRESCO \citep{FRESCO}, PRIMER \citep{primer}, NGDEEP \citep{NGDEEP}, EIGER \citep{EIGER}, ASPIRE \citep{ASPIRE}, CEERS \citep{CEERS}, PEARLS \citep{PEARLS}, GLASS-JWST \citep{GLASS}, etc., are accumulating sky area and sample statistics to enable robust measurements of the demography and properties of the first galaxies and SMBHs. With demonstrated instrument capabilities and survey strategies from these JWST programs, the time is ripe to carry out ambitious surveys that will ultimately answer key questions pertaining to the cosmic emergence of the first galaxies and SMBHs. When carefully designed, such a survey that combines unique capabilities of JWST in terms of wavelength and depth can transform multiple science areas.

\begin{figure*}[]
\centering
\includegraphics[width=0.95\textwidth]{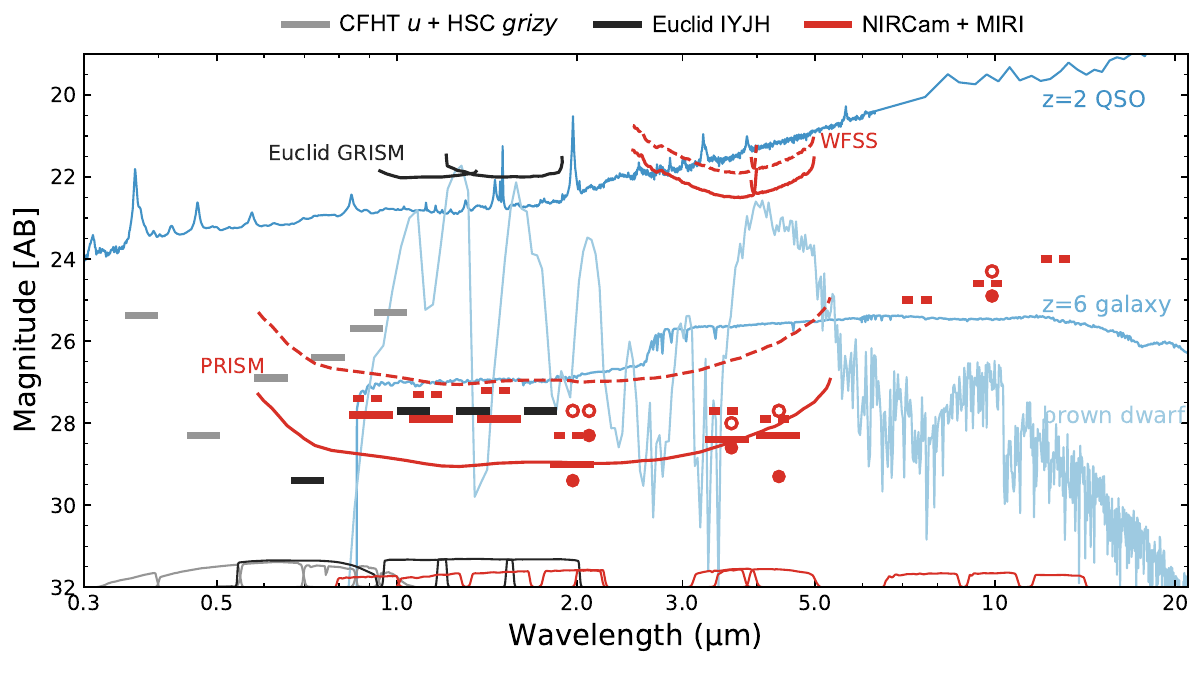}
\caption{Continuum sensitivity (5$\sigma$ for imaging, ${\rm S/N/pixel}=3$ for spectroscopy; point source) and wavelength coverage of different surveys. The \proj\ JWST observations are highlighted in red. Short horizontal segments represent photometry and curved lines represent spectroscopy. Dashed and solid format refers to epoch and final depths, with the open/filled circles for the Deep tier. Depth details are in {Table~\ref{tab:depth}}. In addition to the deep public optical imaging \citep{heroes} from Subaru HSC and 0.9--2\,\micron\ spectroscopy from Euclid, this field is fully covered by deep multi-wavelength data from radio to X-rays. Three representative templates for a bright $z=2$ broad-line quasar, a $z=6$ passive galaxy and a brown dwarf are shown in blue. Note that emission-line-dominated galaxies can be detected in NIRCam/WFSS and Euclid/grism to much fainter continuum depths than shown here. }
\label{fig:depth}
%\hspace{-2cm}
\end{figure*}

\proj\ (PID: 5105) curates a multi-cycle JWST imaging and spectroscopic survey to address three fundamental thrusts: {\bf (1)} galaxy science with a focus on the $z\gtrsim 6$ populations; {\bf (2)} time-domain science focusing on transients and AGN reverberation mapping; {\bf (3)} data challenges and engagements for the broader JWST and astronomical communities. It consists of two tiers: a {\bf Wide} tier covering $\sim 400\,{\rm arcmin}^2$ with 2.4--5\,\micron\, NIRCam/WFSS grism spectroscopy, and a {\bf Deep} tier covering the central $\sim 50\,{\rm arcmin^2}$ with 0.6--5.3\,\micron\, NIRSpec MOS/PRISM spectroscopy. Both tiers have NIRCam and parallel MIRI imaging covering $\sim 0.9-12$\,\micron, though the Deep tier has fewer filters. \proj\ is by far the most ambitious deep spectroscopic GO program with JWST, but also carries with a large amount of ``free'' deep JWST imaging time to improve survey efficiency.

The time-domain element is a unique aspect of \proj, and naturally flows down from the depth requirement. The Wide tier will have three epochs and the Deep tier will have 18 epochs, all distributed over 3 cycles (a total baseline of 3+ yrs). These timing constraints restrict the field to be near the ecliptic poles. Since our primary focus is on deep extragalactic science, the South Ecliptic Pole region is less favorable due to its proximity to the LMC. We choose a North Ecliptic Pole (NEP) field that not only provides unlimited year-round visibility for JWST, but also stable, low background (dominated by the zodiacal light). More importantly, the \proj\ field is within the continuous view zone of many space missions, e.g., eROSITA \citep{Merloni_etal_2012}, WISE \citep{Wright_etal_2010}, Roman \citep{WangY_etal_2022}, etc., with incredibly rich and deep multi-wavelength data that are either existing or coming soon. In particular, the \proj\ field is fully within the Euclid Deep Field North, with deep optical-NIR photometry and grism spectroscopy up to 2\,\micron. Figure~\ref{fig:depth} summarizes the major multi-wavelength data covering this field. \proj\ observations greatly expand the wavelength coverage and elevate the legacy value of all these multi-wavelength data.  

As a treasury, \proj\ is optimally designed to maximize the return in several key science areas in one go, and the driving science goals cannot be achieved with existing programs. Specifically, it has roughly the second largest contiguous area next to COSMOS-Web with comparable imaging depths, yet with massive spectroscopy and more filters. It will provide {one of the first} high-quality measurements of galaxy clustering at $z\gtrsim 6$ to connect galaxies to dark matter halos \citep[e.g.,][]{trinity}. Unlike most high-$z$ WFSS programs with only one filter, such as the Cycle-3 Treasury program COSMOS-3D, the maximum IR spectral coverage by \proj\ enables detailed studies not only for galaxies at Cosmic Dawn, but also for $1<z<6$ galaxies, reaching several magnitudes fainter than past studies. The deep NIRSpec/PRISM spectroscopy is motivated by the success of programs such as UNCOVER \citep{UNCOVER} and JADES \citep{JADES}, but \proj\ will achieve uniform targeting for a much larger, flux-limited sample. 

In addition to galaxy and AGN science, \proj\ will perform the first systematic and transformative time-domain study with JWST. Unlike other heavily-invested extragalactic fields with limited annual visibility (such as GOODS and COSMOS fields), the choice of a NEP field ensures cadenced visits, and the unprecedented imaging depths of JWST will greatly facilitate the discovery and characterization of high-redshift (e.g., $z>3$) transients and variables. The 18 Deep epochs and the large amount of available NIRSpec/MSA slots will provide further utility for deep spectroscopic follow-up of these high-$z$ transients. 

While \proj\ spectroscopy does not go as deep as some of the deepest spectroscopic targets in, e.g., JADES and UNCOVER, it will provide a much more complete, and a much more massive ($\sim 10^4$) spectroscopic sample with uniform targeting. The overall design of \proj\ ensures maximum synergy across different science categories, and has the potential to revolutionize these areas. The combination of sky area, depth, spectral coverage and statistics, and time-domain aspects is the greatest strength of \proj. 

\begin{figure}[]
\centering
\includegraphics[width=0.48\textwidth]{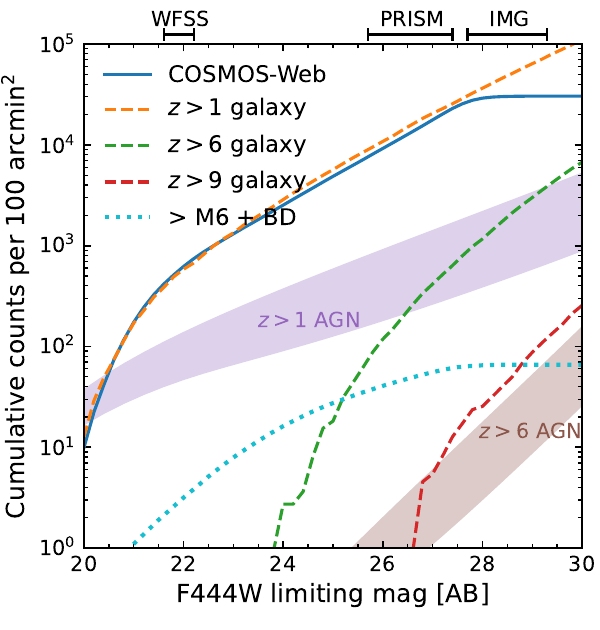}
\caption{Photometric source counts as a function of continuum limiting magnitude at F444W (details in Table~\ref{tab:counts}). Predicted total galaxy counts \citep{JAGULAR_mock} are roughly in line with the observed counts in COSMOS-Web \citep{COSMOS-Web,Zhuang_etal_2024a}; the latter is flux-limited at $\sim 28$ mag in F444W. Actual counts for $z\gtrsim 6$ AGNs at these faint magnitudes may be significantly higher given recent JWST observations \citep{YangG2023}. The ranges of continuum depth reached by different observing modes of \proj\ are marked near the top. }
\label{fig:counts}
%\hspace{-2cm}
\end{figure}

To summarize in numbers, \proj\ will produce: $>10^5$ photometric galaxies with 0.9--12\,\micron\ coverage, $\sim 2500-3800$ galaxies (F444W$\lesssim 22.2$) with 2.4--5\,\micron\ $R\sim 1400$ grism spectra, $\sim 5000-10^4$ galaxies (F444W$\lesssim 26$) with 0.6--5.3\,\micron\ low-$R$ PRISM spectra, $\gtrsim 1000$ spectroscopic $z> 5$ galaxies, $\sim 50$ SNe Ia and $\sim 170$ core-collapse SNe at $z>3$, and $\sim 300$ brown dwarfs (see Table~\ref{tab:depth} and Table~\ref{tab:counts}). These sample statistics are based on the expected source counts of various populations both from simulations and real observations (shown in Fig.~\ref{fig:counts}), as detailed in \S\ref{sec:sci}. 

\begin{figure}
%\begin{figure}
%\epsscale{0.85}
 %\plotone{sample.eps}
 \includegraphics[width=0.48\textwidth]{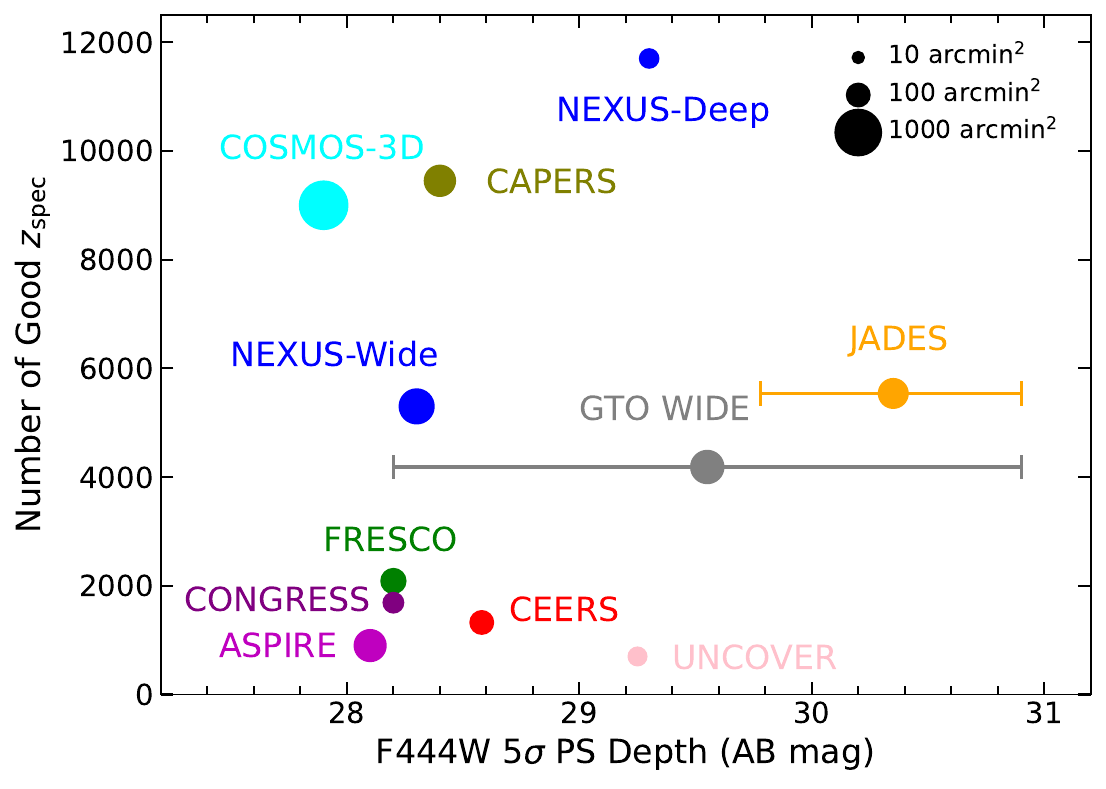}
 \caption{Comparison of large JWST spectroscopic programs in terms of the yield of reliable spectroscopic redshifts. For \proj, the spec-z yields correspond to the target limiting magnitude that will result in a spectral continuum S/N$>3$ per pixel, regardless of spectroscopic instruments ( NIRSpec/MOS or NIRCam/WFSS). For other programs, we adopt the nominal forecast numbers or available MSA slots in these programs. This spec-z yield is plotted against the imaging depth of each survey, though the spectroscopic subset of targets are generally much brighter than this imaging depth. The size of the symbol is linearly proportional to the area of the survey. {ASPIRE imaging depth is for F356W. As CONGRESS has the same sky coverage in GOODS-N as FRESCO, we adopt the same F444W imaging depth as FRESCO.}} \label{fig:comp}
%\end{figure}
\end{figure}

To further place the \proj\ spectroscopic sample in context, we compare the expected number of objects with reliable spectroscopic redshifts\footnote{Our definition of reliable spectroscopic redshifts is that the spectral continuum signal-to-noise ratio per pixel exceeds 3, and the relative numbers of good spectroscopic redshifts from different programs are scaled according to their area, depth, and expected source counts versus limiting magnitude. This is a more stringent definition than all targets that will have a spectrum from NIRCam/WFSS or NIRSpec/MOS. In practice, it is possible to derive a spectroscopic redshift even if the continuum S/N is as low as 1.} for several large JWST programs in Fig.~\ref{fig:comp}, as a function of imaging depth in NIRCam F444W. These redshift surveys include JADES, the NIRSpec Wide GTO Survey \citep[GTO-WIDE;][]{GTO-WIDE}, CEERS (ERS 1345, PI: Steven Finkelstein), UNCOVER, CAPERS (GO 6368; PI: Mark Dickinson), COSMOS-3D (GO 5893; PI: Koki Kakiichi), FRESCO \citep{FRESCO}, CONGRESS (GO 3577, PI: Eiichi Egami), and ASPIRE (GO 2078; PI: Feige Wang). The combination of the three JWST Cycle-3 Treasury programs (COSMOS-3D, CAPERS and NEXUS), along with pure-parallel grism spectroscopic programs such as SAPPHIRES (GO 6434), will deliver the largest number of spectroscopic redshifts from JWST. 

This overview paper describes the \proj\ project. In \S\ref{sec:sci} we elaborate on the science rationale for \proj, with highlights on key science cases. In \S\ref{sec:design} we describe the program design and implementation. In \S\ref{sec:synergy} we describe the data production roadmap and community opportunities. We summarize in \S\ref{sec:sum}. Throughout this paper, we adopt a flat $\Lambda$CDM cosmology with $\Omega_\Lambda=0.7$ and $H_0=70\,{\rm km\,s^{-1}Mpc^{-1}}$.

\section{Science}\label{sec:sci}

\subsection{Galaxies and Supermassive Black Holes}\label{sec:gal}

\proj\ aims to assemble large, uniform, deep photometric and spectroscopic samples to study the evolution of galaxies and SMBHs from Cosmic Dawn to Dusk. \proj\ will measure, for the first time, the large-scale clustering of galaxies at $z\gtrsim 6$ with spectroscopic samples. Such measurements are critical to understanding galaxy formation at Cosmic Dawn, and are only possible with JWST. This key science goal necessitates a {\bf Wide} tier, with area of hundreds of arcmin$^2$ to probe projected comoving scales of tens of $h^{-1}{\rm Mpc}$. Grism spectroscopy with $R>1000$ can efficiently identify line-emitting galaxies at $z>6$ over this entire area \citep[e.g.,][]{FRESCO,EIGER, ASPIRE,Sun_2023}, reaching an equivalent broad-band magnitude of F444W$\sim 27.5$. This depth is desirable to reach the number density and galaxy mass of interest for clustering measurements (see below). 

However, WFSS is insensitive to continuum-dominated sources: the limiting magnitude is only $\sim 22$ mag to reach a continuum S/N$>3$ per pixel from WFSS. While this does not impact clustering statistics (dominated by line emitters), to fully explore a diverse $z>6$ galaxy population, a large spectroscopic sample to much fainter magnitude is necessary. For example, the recently discovered two $z\sim 14$ galaxies from JADES \citep{JADES_z14} are not dominated by line emission, but rather continuum emission from a well-established stellar population, thus would have eluded discovery with WFSS. One goal of \proj\ is to identify these earliest galaxies with nearly complete spectroscopy. It is prohibitive to cover the entire Wide tier with deep spectroscopy. The solution is to focus on a much smaller area and perform high-throughput NIRSpec PRISM spectroscopy for a large sample. The {\bf Deep} tier, with 18 epochs to significantly boost multiplexing, will provide up to $\sim 10^4$ spectroscopic redshifts for a nearly complete photometric sample to $m_{\rm AB}\sim 26$ (at 4~\micron), with hundreds at $m_{\rm AB}\sim 27-28$ coadd depth (continuum dominated). This will provide a benchmark sample with $0.6-5.3$~\micron\ spectra to study physical properties of galaxies. The combination of Wide and Deep ensures the most complete spectroscopic sample at $z>6$, but also maximizes the utility of deep photometry and spectroscopy for galaxies at all redshifts.

\subsubsection{Galaxies and AGNs at Cosmic Dawn: Abundance, Clustering, and Physics}

\noindent\fire{Demography of the first galaxies and SMBHs}\quad \proj\ will reach F444W$\sim 28$~AB in imaging over a $0.1\,{\rm deg^2}$ field to provide $>4000$ photometric galaxies at $z>6$ (dominated by star-forming galaxies, SFGs). The Wide+Deep tiers will deliver IR spectra for $\gtrsim 25\%$ of them (Table~\ref{tab:counts}). Furthermore, the NIRCam/grism and NIRSpec/PRISM spectra can identify AGNs among these galaxies, assisted by UV-MIR imaging, deep X-ray and long-term variability. This would produce by far the largest, and the most complete, spectroscopic AGN sample at $z>6$ to robustly measure their abundance, clustering, and physical properties.

AGN identification at $z>6$ will take advantage of the broad WFSS grism wavelength coverage in Wide and sensitive PRISM spectra in Deep, assisted by deep multi-band photometry and variability. Type 1  (unobscured/broad-line) AGNs can be directly identified by the presence of broad ($\rm FWHM \gtrsim 1000\ km\ s^{-1}$) permitted lines in the spectra \citep[e.g.,][]{Harikane2023_type1, Maiolino2023_type1, Greene2024_LRD,Matthee2024_LRD}. On the other hand, it is notoriously difficult to robustly identify Type 2 (obscured/narrow-line) AGNs at $z\gtrsim 3$, as the commonly used diagnostic diagrams at low redshift do not work for these high-$z$ AGNs due to the global change of metallicity and ionization conditions \citep[e.g.,][]{Harikane2023_type1, Maiolino2023_type1, Scholtz2023_type2}. The presence of high-ionization lines in combination with additional UV/optical transitions (e.g., \NeV$\lambda3420$, \OIII$\lambda4363$, \HeII$\lambda4686$) aided by photoionization modeling provides a promising means for identifying high-redshift narrow-line AGNs in the low metallicity regime \citep[e.g.,][]{Scholtz2023_type2, Mazzolari2024}. The MIRI data will also facilitate the identification of heavily obscured AGNs by capturing the reprocessed hot dust emissions from the AGN torus \citep[e.g.,][]{YangG2023, Lyu2024_MIRI, PerezGonzales2024_MIRI}. 

At the design depths, we are probing BH masses down to $\sim 10^{6-7}\,M_\odot$, meaning these AGNs are the growing progenitors of the $10^9\,M_\odot$ SMBHs observed in $z>6$ quasars. The clustering measurements constrain their host halo masses, and thus will provide critical insights on their seeding scenarios (e.g., are they in massive halos formed earlier, or low-mass SFG hosting halos). Spectroscopy (w/ broad-line detection) further provides single-epoch BH mass measurements \citep{Vestergaard_Peterson_2006,Shen_2013} and thus Eddington ratios for these accreting SMBHs, offering additional constraints on this rapid growth period. Thanks to the well-defined survey selection function and the nearly complete spectroscopic sample (down to $m_{\rm AB} \sim 26$ mag in F444W in deep), we will be able to provide the most robust constraints on the faint end of the AGN luminosity function and the active BH mass function to facilitate comparisons with theoretical predictions \citep[e.g.,][]{Trinca2022_BHMF, Li2024_BHMF}.  Put together, we are closing in on the key question of the formation of billion-solar-mass SMBHs at Cosmic Dawn and the SMBH seeding mechanisms. 

An emerging theme from recent JWST observations is the discovery of compact, red, high-redshift galaxies with prevalent AGN signatures (i.e., broad Balmer lines), aka, the ``little red dots'' \citep[LRDs, e.g.,][]{Labbe2023, Matthee2024_LRD, Greene2024_LRD, Kokorev2024_LRD, Akins2024_LRD, Kocevski2024_LRD, Wang2024_LRD}. While the non-broad-line LRDs may be dusty star-forming galaxies \citep[e.g.,][]{PerezGonzales2024_MIRI}, the non-detection of broad-line LRDs in deep stacked X-ray/radio/sub-mm data has raised significant debate about their AGN nature \citep[e.g.,][]{Ananna2024_LRD, Yue2024_LRD, Akins2024_LRD, Juodvzbalis2024_CTAGN, Maiolino2024_CTAGN}. The unique time domain aspect of \proj, especially in the deep tier, enables us to construct the most well-sampled long-term light curves of LRDs to reveal their stellar vs. AGN nature through variability signatures \citep[e.g.,][]{Kokubo2024_LRD}. The comparison of rest-frame UV/optical emission lines and SED properties of LRDs with the general AGN population identified in \proj\ will further shed light on their nature.

\begin{figure}
%\begin{figure}
\epsscale{0.85}
 %\plotone{sample.eps}
 \includegraphics[width=0.48\textwidth]{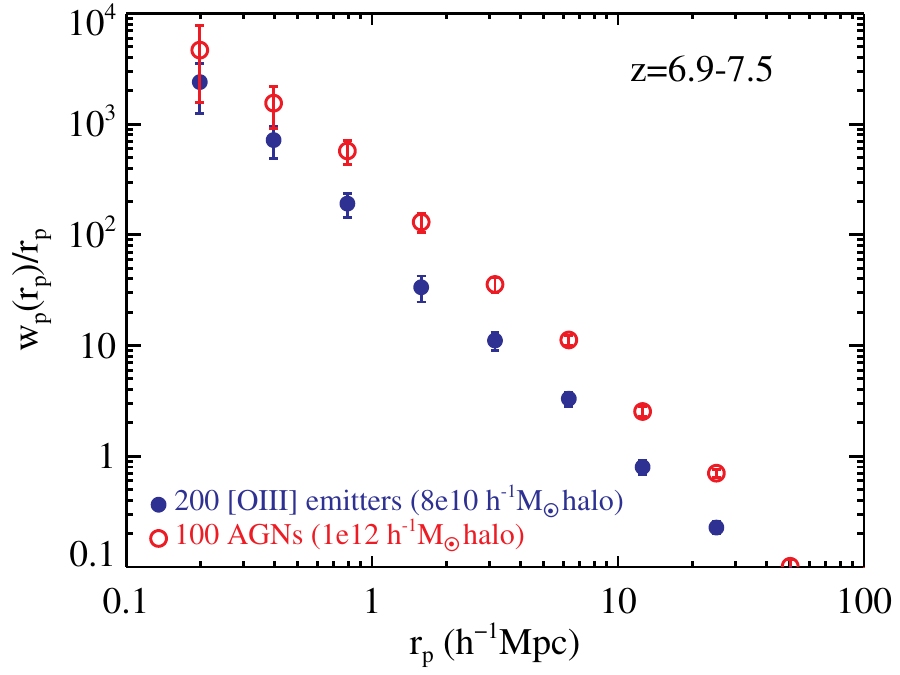}
 \caption{Predicted clustering (projected correlation function $w_p(r_p)$) measurements for spectroscopic WFSS sources at $z\sim 6.9-7.5$. For simplicity we have assumed a power-law real-space correlation function $\xi(r)=(r/r_0)^{-1.8}$, where the correlation length $r_0$ corresponds to a given host halo mass in this redshift range, assuming linear bias. The anticipated spectroscopic sample from the Wide tier is able to distinguish different halo mass scales with clustering measurements to constrain SMBH seeding scenarios.} \label{fig:clustering}
%\end{figure}
\end{figure}

\noindent\fire{Clustering of galaxies and AGNs}\quad For clustering measurements, the pair count at a given projected scale (e.g., $5~h^{-1}{\rm Mpc}$) is $N_{\rm pair}\propto n^2A$, where $n$ and $A$ are the sky density and survey area. At $z>6$, the galaxy counts roughly double for every 0.5 mag increase \citep{JAGULAR_mock}. Thus given fixed observing time, doubling the exposure time is better than doubling the area, as long as the total area is sufficient to measure the correlation function on linear scales (tens of $h^{-1}{\rm Mpc}$). Recent line-emitter searches using FRESCO WFSS/grism data suggest $\sim 200$ \OIII\ emitters at $6.7\lesssim z\lesssim 8$ and $\sim 700-800$ \halpha\ emitters at $z=4.9-6.6$ over \proj\ Wide (F. Sun, private communications). Figure~\ref{fig:clustering} forecasts the clustering measurements at $z\sim 7.2$, based on the WFSS/grism sample alone. For typical $8\times 10^{10}\,h^{-1}M_\odot$ halos of these \OIII\ emitters, the correlation function (blue filled circles) can be measured with high precision. The red open circles represent a scenario of 100 growing SMBHs or AGNs in $\sim 10^{12}\,h^{-1}M_\odot$ halos that are more typical of luminous quasars at $z\gtrsim 6$ \citep{Arita2023}. Using the much larger \halpha-emitter sample or the full photometric sample at $z\gtrsim 6$ would further improve the clustering measurements at Cosmic Dawn. 

\noindent\fire{Physical properties of galaxies}\quad The grism and prism spectra over broad wavelength ranges will be essential for probing galaxy physics at
$z>6$. NIRCam/WFSS $2.4-5$\,$\mu$m spectroscopy over Wide will cover key rest-optical emission lines in this redshift
range, including \OII$\lambda3727$, H$\beta$, \OIII$\lambda\lambda4959,5007$, and H$\alpha$. NIRSpec PRISM spectroscopy in Deep
will include not only these features but also rest-UV lines such as Ly$\alpha$, \CIV$\lambda1549$, \HeII$\lambda1640$, and \CIII$\lambda 1908$. This suite of emission lines provides a rich window into the astrophysics of star formation and chemical enrichment in the ISM.
Additionally, the combined analysis of Ly$\alpha$ and rest-optical emission lines at $z>7$ can lend insights into the
nature of ionized bubbles during the process of cosmic reionization.

A basic probe of excitation and ionization in SFGs is the equivalent width (EW) of \OIII+H$\beta$ emission lines, which is sensitive to galaxy specific SFR and metallicity. Based on photometry, \citet{Endsley2021} and \citet{Endsley2023} have quantified the distribution of
EW(\OIII+H$\beta$) at $z\sim 6.5-8$, showing that there is only weak luminosity dependence if any over the range $0.4-2 L*$, and that, furthermore,
$\sim 20$\% of $z>6.5$ faint SFGs show both high ($>20$~Gyr$^{-1}$) sSFR and low ($<600$\AA) EW(\OIII+H$\beta$). Such systems
may be indicative of low metallicity, high escape fractions of ionizing radiation, or time-variable star-formation histories. Spectroscopic
measurements of EW(\OIII+H$\beta$), along with stellar population analysis, are critical for understanding the ionizing properties
of $z>6$ SFGs.

Analyzing \OII, H$\beta$, \OIII, and H$\alpha$ in concert (and \NeIII$\lambda 3870$ if available) yields powerful constraints
on the oxygen abundance in SFGs. The spectra collected here will cover most, if not all, of these features for galaxies at $z>6$,
and can be used to trace the evolution of mass-metallicity relation \citep[MZR;][]{Tremonti2004} and fundamental metallicity relation \citep[FMR;][]{Mannucci2010}
during the epoch of reionization. Preliminary results with JWST within this redshift range are based on small samples and/or very limited sets of emission lines \citep{Curti2023,Nakajima2023,Matthee2023}. Based on the expected 100 $z>6$ galaxies in Deep that have sensitive rest-optical
spectra and stellar population measures (mass, SFR), it will be possible to trace the star-forming galaxy metallicity scaling relations with high confidence.
Quantifying the normalization, slope, and scatter of the MZR and FMR and comparing with measurements at lower redshift \citep{Sanders2021}, it will be possible to distinguish among different models for galaxy formation feedback \citep{Langan2020}.

NIRSpec PRISM spectra in Deep will enable a joint analysis of rest-UV and rest-optical features for galaxies at $z>6$. As shown in \citet{Tang2023}, the strength of rest-UV Ly$\alpha$ emission at fixed rest-optical emission-line properties (e.g., \OIII/\OII\ ratio) appears
to be depressed at $z>7$, relative to lower redshifts. Such a deficit can be explained if galaxies are surrounded by relatively small
ionized bubbles. However, the sample in \citet{Tang2023} is only 21 galaxies. With \proj, we will have rest-UV
plus rest-optical spectra for a sample 5 times larger than in previous work, enabling a statistical study of the escape
of Ly$\alpha$ emission.

\subsubsection{Galaxy and AGN Science at $1<z<6$}

\proj\ will assemble a complete, magnitude-limited sample of $\sim 5000$ galaxies down to $m_{\rm AB,4\micron}\sim 26$ with $0.6-5.3$~$\mu$m PRISM spectra in Deep.
This sample will be dominated by SFGs at ``Cosmic Noon" (i.e., $z\sim 1-3$). The value of rest-optical
spectroscopy for galaxies at Cosmic Noon has been demonstrated using ground-based surveys such as MOSDEF
and KBSS \citep{Kriek2015,Steidel2014}. A vast complement of science goals can be achieved based on measurements
of multiple strong rest-optical nebular lines in concert with stellar population properties (e.g., stellar mass, size, SFR).
The key scaling relations of the star-forming galaxy population have been constructed, including the galaxy star-forming main sequence \citep{Shivaei2015}, the mass-metallicity and fundamental metallicity relations \citep{Strom2017,Sanders2021}, the scaling of dust attenuation with mass and SFR \citep{Whitaker2017,Shapley2022}, and the emission-line excitation sequences \citep{Shapley2015,Shapley2019}. However, our spectroscopic sample
will be an order of magnitude larger {and deeper} than these previous studies and therefore shed light on the intrinsic scatter in such relations, essential for understanding their origin and evolution. 

Similar to the case for $z>6$ galaxies, \proj\ will efficiently identify the most complete $1<z<6$ AGN sample (both unobscured and obscured) with the broad PRISM spectral coverage, deep optical-MIR photometry, variability, and ancillary multi-wavelength data. This unprecedented complete spectroscopic sample with JWST imaging will revolutionize studies on SMBH-galaxy scaling relations, galaxy mergers, dual AGNs, galaxy groups and proto-clusters, and gravitational lenses across cosmic time \citep[e.g.,][]{Duncan2019, Li2024_dual, Li2024_mass, Tanaka2024}. 

\subsubsection{Brown Dwarfs (and the Milky Way as a Galaxy)} 

The coldest stars and brown dwarfs \citep[late-M, L, T and Y dwarfs;][]{2005ARA&A..43..195K} are the second-most abundant objects in ultra-deep 2-5\,\micron\ sky ($<1\%$ of all galaxies), and are important probes of low-mass star formation over the full history of the Milky Way \citep{2009IAUS..258..317B,2016AJ....151...92R}.  \proj\ will identify and characterize hundreds of brown dwarfs out to kiloparsec distances with multi-band photometry, multi-epoch astrometry (proper motions) and grism/prism spectroscopy. 
This unprecedented sample will enable a statistically robust measurement of the age-dependent thin disk scale-height down to the 500~K T/Y dwarf boundary, probe brown dwarf formation in the ancient thick disk and halo, and provide empirical tests of metallicity-dependent evolutionary and atmosphere models.

Our NIRSpec PRISM spectra will enable exploration of temperature, gravity, metallicity, non-equilibrium mixing, and cloud properties for a diverse sample of brown dwarfs as probed by H$_2$, H$_2$O, CH$_4$, CO, CO$_2$ and PH$_3$ molecular features (Figure~\ref{fig:depth}), the latter recently recognized as a potential metallicity diagnostic \citep{2023arXiv230812107B}. With the expectation of identifying dozens of thick disk dwarfs with multi-band photometry, spectroscopy, and kinematics, we will make the first robust measurement of the thick/thin ratio in the substellar regime, a test of metal dependence on low-mass star formation \citep{2022ApJ...934...73A}. The addition of MIRI imaging and associated broad-band colors will also provide diagnostics of H$_2$O, CH$_4$ and NH$_3$ absorption that trace chemistry and non-equilibrium mixing, and silicate absorption that traces cloud formation and variability, the latter facilitated by our multi-epoch measurements.

We forecast the expected numbers of brown dwarfs in the \proj\ NEP field using the SDSS Milky Way structure model \citep{2008ApJ...673..864J}, state-of-the-art brown dwarf evolutionary models \citep{2021ApJ...920...85M}, and empirical relations to map present-day physical parameters to JWST observables \citep{2023arXiv230812107B}. The \proj\ field points at a lower Galactic latitude than prior deep JWST surveys ($b$ = 30$^o$), which results in a higher surface density of thin and thick disk brown dwarfs. We estimate $\sim 300$ late-M, L and T dwarfs will be detected in at least three bands in the Wide survey, including $\sim$15 thick disk and halo L and T subdwarfs. The Deep survey will detect $\sim$40 late-M, L, T, and Y dwarfs in two imaging bands and with NIRSpec/PRISM spectroscopy (Table~\ref{tab:depth}), with a sample dominated by T dwarfs probing multiple Galactic populations. These samples will be 5--10$\times$ larger than those so far uncovered in other deep JWST fields \citep{2023arXiv230812107B,2023arXiv230903250H,2023arXiv230810900L,Hainline24_BD}, and will enable statistically robust studies of the spatial distributions of brown dwarf populations and the influence of metallicity on low-temperature atmospheres.

\subsection{Time Domain}\label{sec:td}

Depth requirements for galaxy science mandate multiple visits to the same pointings, enabling cutting-edge time-domain science with negligible ($\sim 2.5\%$) initial slew overhead. In Wide, \proj\ provides 3 visits (limited by tiling overhead) separated on yearly timescales; in Deep, a $\sim$2-month cadence optimal for transient searches and AGN reverberation mapping can be accommodated. A maximum baseline of 3 years (cycles 3--5) is optimal for SN searches, AGN variability and proper motion measurements (for brown dwarfs). The 18 epochs of the Deep tier further allows a large number of NIRSpec/PRISM target slots, significantly expanding the multiplex capability of NIRSpec (i.e., $\sim 10^4$ slots for targets as faint as F444W$=27.4$; see Table~\ref{tab:depth} and \S\ref{sec:design}). 

\begin{figure*}[]
\centering
\includegraphics[width=0.49\textwidth]{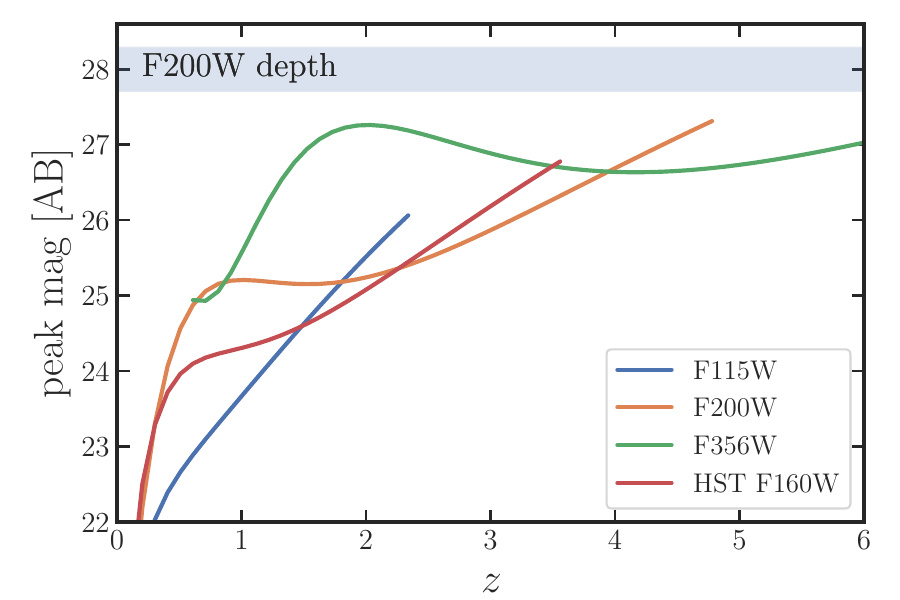}
\includegraphics[width=0.49\textwidth]{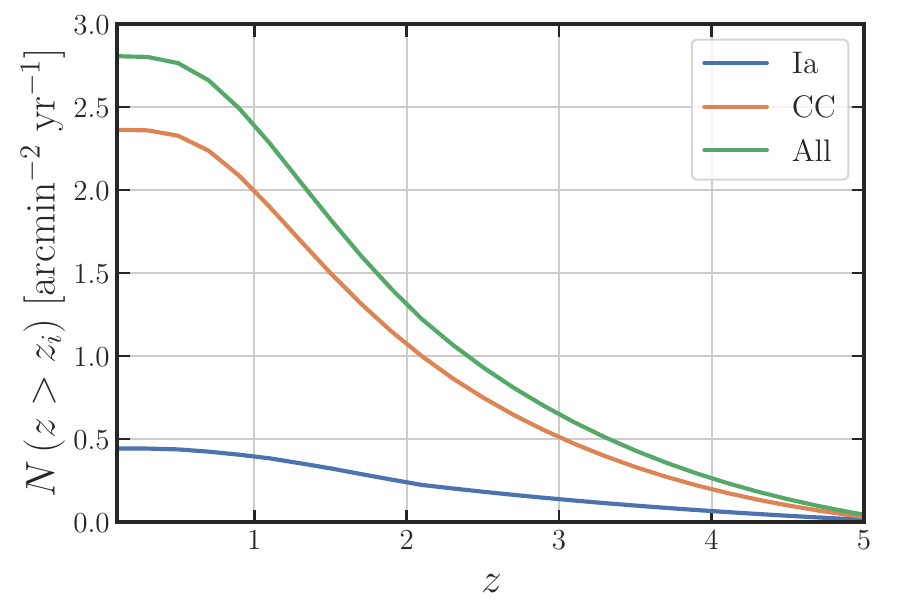}

%\hspace{-1cm}
\caption{SNe forecasts. \textbf{Left:}  Peak magnitudes as a function of redshift for a typical SN Ia with $M_V = -19.5$ mag, using a polynomial fit to the SN Ia template in \cite{Lu2023_SNIa} at $\lambda_{\rm rest} \sim 0.4-2\ \mu m$. Many classes of CC SNe (not shown) are $\sim 1$~mag fainter. At $z\gtrsim 3$, F200W traces the rest-5000\,\AA\ and is optimal for high-$z$ SN searches. Our single-epoch imaging depths (shaded gray) are sufficient to detect these SNe even when they fade by $\sim 2$~mag from peak. \textbf{Right:} Predicted \textbf{annual number} of SNe per 1~${\rm arcmin^2}$ area. For Ia, the volumetric rate from CANDELS \citep{Rodney_Ia_rate} is adopted. CC SNe rate is from \citet{StrolgerCCRate}. At $z\gtrsim 3$, two observed-frame months are $\lesssim 15$ rest-frame days when the SN remains detectable by \proj\ NIRCam imaging. Thus the SNe search efficiency is nearly 100\% in Deep, and $\sim 2\,{\rm mo}/12\,{\rm mo}=16.7\%$ in Wide. Predicted numbers of detectable $z>3$ SNe Ia are listed in Table~\ref{tab:counts}. }
\label{fig:sn}
\end{figure*}

\subsubsection{Supernovae}\label{sec:sne}

While discoveries of high-redshift transients with multi-epoch JWST imaging have been made possible with other programs \citep[e.g.,][]{Yan_2023, OBrien_etal_2024,jades_sn}, \proj\ will perform the first large-scale systematic and efficient SN searches at $z\gtrsim 3$ using the optimal F200W filter, to robustly quantify their rates and characterize their physical properties. Given the area coverage and cadences, we estimate that \proj\ will discover $\sim 50$ SNe~Ia and $\sim 170$ core-collapse (CC) SNe at $z>3$ (Fig.~\ref{fig:sn} and Table~\ref{tab:counts}). The overall expected numbers of SNe are roughly in line with the observed rates in other JWST multi-epoch imaging programs \citep[e.g.,][]{Yan_2023, OBrien_etal_2024,jades_sn}. These SN samples will robustly measure their rates at $z\gtrsim 3$ for {the first time} \citep{Rodney_Ia_rate,StrolgerCCRate}. At these redshifts, time-dilation allows us to obtain multiple exposures for each SN in Deep, along with deep PRISM spectra (Fig.~\ref{fig:depth}) for reliable classification. In Wide, SN classification will largely rely on host galaxy properties. The large number of spec-zs and reliable photo-zs in \proj\ are ideal for high-redshift SNe studies. Each $z\gtrsim 3$ SN provides an invaluable constraint. 

SNe~Ia at $z\gtrsim 3$ can constrain any evolution in the dark energy equation-of-state, $w(z)$ \citep{Riess2018, Lu2022}, and this dataset will be complementary to the SNe~Ia sample that the Vera C. Rubin Observatory will obtain in its deep drilling fields \citep{Gris2023}. Measuring these high-redshift SNe~Ia rates and properties is invaluable to tuning the survey strategy for the Nancy Grace Roman Space Telescope's (Roman) High Latitude Time-Domain Survey. Moreover, all SNe (SNe~Ia or CC SNe) occurring at these redshifts will be in very different host galaxies from those at low redshift, and this high-redshift sample will allow us to understand the correlation between SN-luminosity, host-galaxy mass and star-formation rate. For SNe~Ia, this can improve their standardizabiliity, while for CC SNe, our program will provide an unprecedented view of some of the earliest stellar deaths, and how they impact cosmic chemistry, together with constraining their cosmic evolution. No ground-based facility will be able to follow these highest redshift SNe, but JWST ToO observations can obtain ancillary observations for them, though the science we describe is possible without these follow-up observations. 

\proj\ will also produce many $z\lesssim 1.5$ SNe. These SNe can be followed up with 200--400s exposures using 2--4m-class telescopes up to $z\sim 0.5$ and with 8m-class facilities over $0.5 \le z \le 1.5$, ensuring we can assemble optical light curves with a few-day cadence to complement the JWST NIR observations. Our ``low-redshift'' SNe Ia sample will be larger than the HST RAISIN 1 \& 2 program \citep{RAISINJones} and SIRAH \citep{Pierel2022} \emph{combined} because of the much higher sensitivity of NIRCam compared to HST/WFC3/IR. As SNe Ia are more standardizable and less impacted by dust in the NIR, this dataset will have tremendous cosmological value. The low-redshift CC SNe sample, while smaller than previous ground-based samples, will enable detailed modeling as the combination of UV-optical-NIR observations strongly constrains their bolometric evolution.  

\subsubsection{High-$z$ Tidal Disruption Events}\label{sec:tde}

\begin{figure*}[]
\centering
\includegraphics[width=0.49\textwidth]{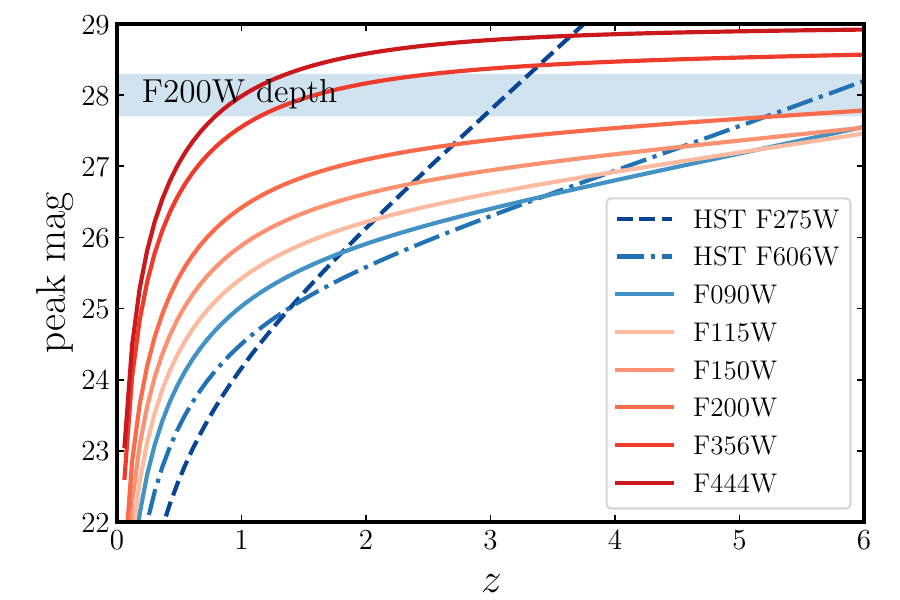}
\includegraphics[width=0.49\textwidth]{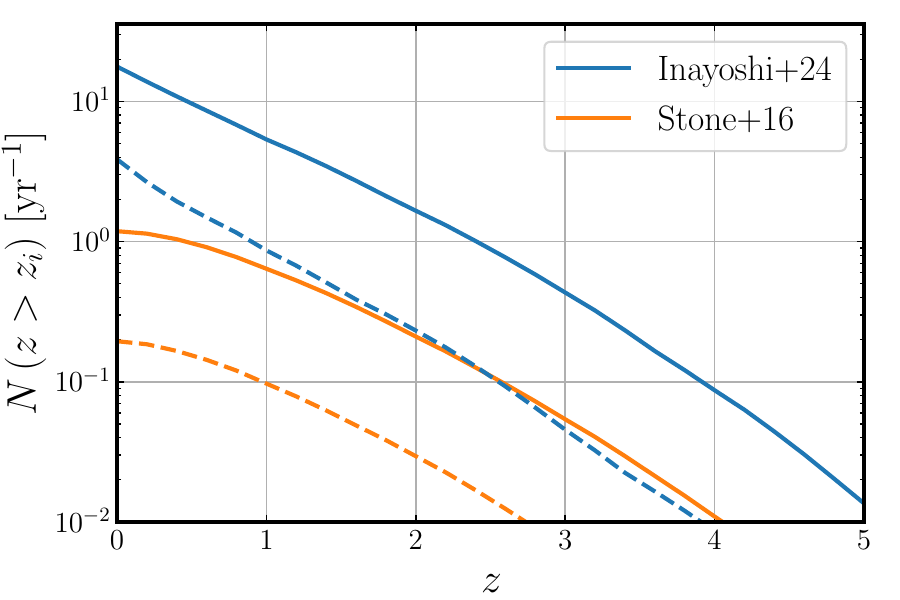}
%\hspace{-1cm}
\caption{TDE forecasts. {\bf Left}: Peak magnitude in various filters as a function of redshift for a TDE with $M_{\rm BH} \sim 10^{5.5}\ M_\odot$ and an Eddington ratio of one, using the SED template of PS1-10jh \citep{vanVelzen2020}. The shaded gray regions indicate our single-epoch imaging depth in F200W, which are sufficient to detect TDEs in such IMBHs. {\bf Right}: Predicted annual numbers of detectable TDEs in the Wide (solid curve) and Deep (dashed curve) \proj\ fields in the F200W band, based on the theoretical TDE rates from \cite{Inayoshi2024} and \cite{Stone2016}. }
\label{fig:tde}
\end{figure*}

When a star passes too close to a SMBH and is disrupted by the BH's tidal forces, about half of its material can be captured and accreted by the BH, producing a bright flare that peaks in the soft X-ray/UV wavelengths \citep{Rees1988}. Since TDEs decay slowly over months to years and can reach the Eddington accretion rate during the rising phase \citep[e.g.,][]{vanVelzen2020, Yao2023}, even an intermediate-mass BH (IMBH) with $M_{\rm BH} \sim 10^{5.5}\ M_\odot$ can be detected as a bright flare in deep single-epoch NIRCam imaging in \proj\ up to $z\sim6$ (Fig.~\ref{fig:tde} left). Therefore, high-redshift TDEs offer a unique opportunity to detect dormant BHs in galaxy centers, and provide the potential to probe lower-mass systems down to the seeding mass regime ($M_{\rm BH} \lesssim 10^6\ M_\odot$) in the early universe, beyond what can be detected in normal AGNs. 

TDEs discovered through ground-based optical campaigns have been largely limited to the low-redshift (e.g., $z<1$) Universe due to the shallow depths of these surveys, with these events surprisingly often found in post-starburst galaxies \citep{French2020}. However, TDEs in star-forming galaxies may be missed due to dust extinction \citep[e.g.,][]{Jiang2021}. \proj, with its deep observations in long-wavelength filters, enables the discovery of dust-obscured TDEs and allows unprecedented detailed studies of their host galaxy properties (e.g., morphology, color) using multi-band NIRCam imaging. In addition, the superb spatial resolution of NIRCam imaging will efficiently pinpoint the location of nuclear transients, facilitating the identification of TDE candidates for follow-up observations. 

The spectroscopic follow-up of these high-redshift TDE candidates can be effectively conducted using NIRSpec, which offers a unique opportunity to probe several key astrophysical processes. Monitoring the spectral evolution of TDEs allows us to study the formation of accretion disks, outflows, and even radio jets in real time to constrain the underlying physics \citep[e.g.,][]{Nicholl2020, Wevers2022, Andreoni2022}. The substantial energy output from TDEs illuminates the ISM near the BHs, providing a rare chance to study the parsec-scale nuclear environments in otherwise inactive galaxies by spectrally resolving the gas echoes photoionized by the flares \citep[e.g.,][]{Wang2012_tde, Charalampopoulos2022}. MIRI follow-up observations of the reprocessed dust echoes will provide crucial insights into the nuclear dust content and distribution \citep[e.g.,][]{Jiang2017}. 

The TDE rates at $z>1$ are unknown, given the lack of observations. We forecast the detectable TDE events in \proj\ using theoretical predictions of the TDE rate at different redshifts. Generally, TDEs are rare phenomena occurring at a rate of $\sim10^{-4}\ \rm yr^{-1}$ per galaxy, with the rate increasing as BH mass decreases. To bracket the large uncertainties in theoretical models, we use an optimistic TDE rate prediction from \cite{Inayoshi2024}, tailored for high-redshift compact systems (assuming $A_V=0.5$ and $R_e=50$ pc for the nuclear star cluster), and a pessimistic model from \cite{Stone2016}. The TDE rate is combined with a model BHMF  \citep{Zhang2023_bhmf} to estimate the number of detectable TDEs, assuming a peak Eddington ratio of unity, and using the SED and light curve for the TDE PS1-10jh \citep{vanVelzen2020}, while accounting for the survey efficiency. Because TDE light curves generally decay more slowly than SNe, the detection efficiency is high even with the low cadence in Wide, in particular for high-redshift TDEs where the $1+z$ time dilation helps. In the optimistic scenario, we expect to detect $\sim5$ TDEs per year at $z>1$, whereas in the pessimistic scenario, $\sim 1-2$ TDEs at $z>1$ are anticipated over the 3-yr survey (Fig. \ref{fig:tde} right). 

This model prediction is severely limited by our poor knowledge of the theoretical TDE rates and the BHMF down to the low-mass end in the high-redshift Universe. The TDE rate in the turbulent, gas-rich high-redshift Universe may be significantly higher than predicted by the classic two-body scattering and relaxation model. This increase could be due to prevalent galaxy mergers, a higher occupation fraction of dense nuclear star clusters in high-$z$ compact galaxies \citep{Pfister2020}, the presence of massive axisymmetric gas disks \citep{Karas2007}, and a higher AGN fraction, where AGN accretion disks may effectively enhance the stellar capture rate \citep{Kaur2024}. 
\proj\ offers currently the best opportunity to systematically search for high-redshift TDEs and constrain their occurrence rates, or at least provide a robust upper limit, which will be crucial for refining theoretical models.

\begin{figure*}[]
\centering
\includegraphics[width=0.48\textwidth]{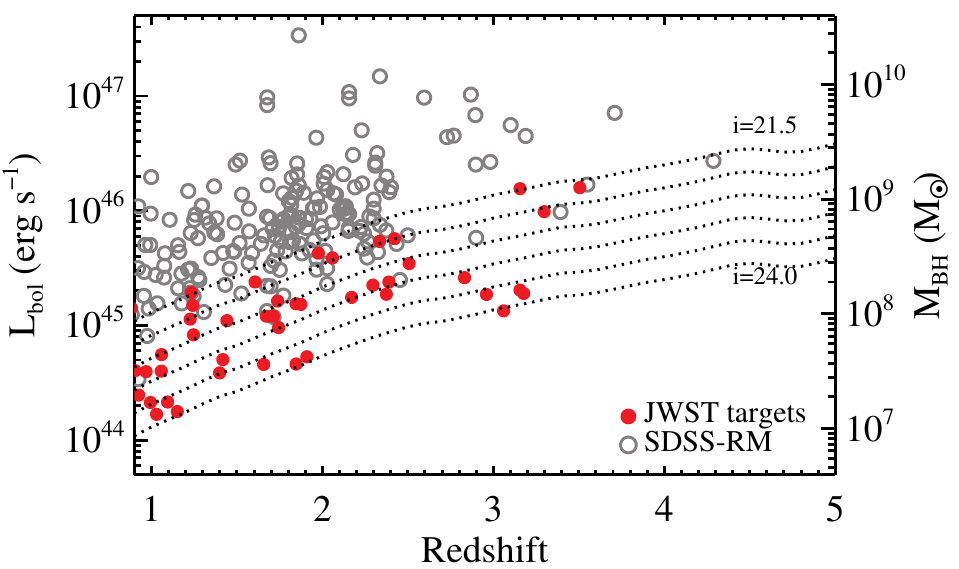}
\includegraphics[width=0.48\textwidth]{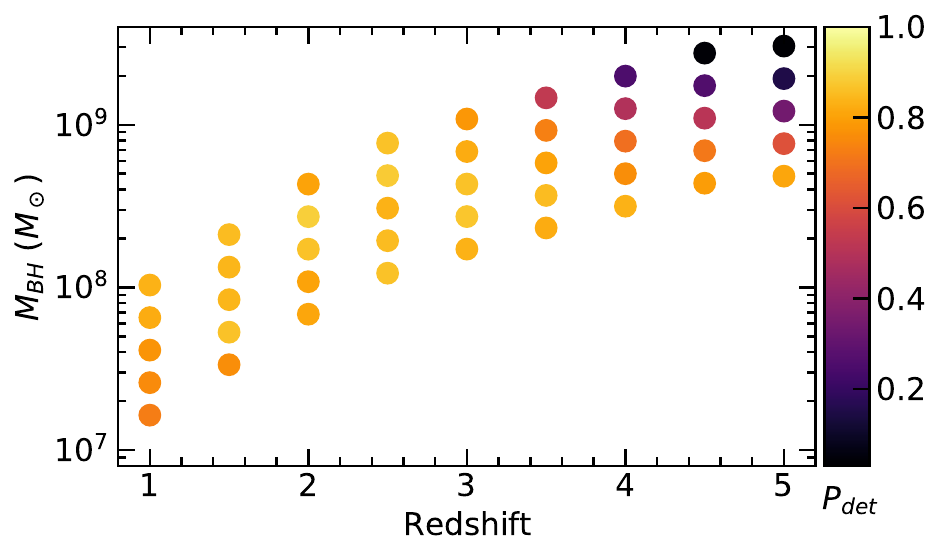}
\caption{AGN RM forecast. {\bf Left:} Sample distribution in the bolometric luminosity-redshift plane. BH mass on the right axis assumes constant Eddington ratio of 0.1. The \proj\ sample is compared with the ground-based SDSS-RM sample \citep{Shen_etal_2023}. {\bf Right:} Detection probability as functions of redshift and BH mass, for the proposed JWST sample. For the vast majority of proposed targets the lag will be detected at high ($\gtrsim 80\%$) probabilities. Lag detection forecast follows earlier approaches \citep{Shen_etal_2015a,King_etal_2015}, with realistic input parameters (AGN variability properties and monitoring parameters). Detection requires $\pm 1\sigma$ from the input lag and an absolute fractional deviation of $<30\%$ of the input. Detection loses sensitivity for $>10^9\,M_\odot$ AGNs at $z>4$ because the expected lags approach the maximum baseline of 3 years. }
\label{fig:rmagn}
%\vspace{-0.5cm}
\end{figure*}

\subsubsection{AGN Reverberation Mapping}\label{sec:rm} 

Our ability of measuring SMBH masses at $z\sim 1-8$ using single-epoch spectra hinges on reverberation mapping (RM) results of low-redshift broad-line AGNs. RM measures a time lag between the driving continuum light curve and the response in the broad-line emission, and this time lag depends primarily on AGN luminosity \citep{Bentz_etal_2013} and potentially also the Eddington ratio \citep{Du_etal_2015,DallaBonta_etal_2020}. However, the situation is unclear for AGNs at $z\gtrsim 2$ if these single-epoch mass recipes based on low-redshift RM AGNs still apply. There is significant concern on the reliability of single-epoch BH mass estimates for the high-redshift LRDs \citep[e.g.,][]{Bertemes_2024}. In order to design robust mass recipes for high-redshift AGNs, it is imperative to obtain RM results covering broad AGN parameter space at $z>1$, most efficiently with multi-object spectroscopic RM programs \citep{Shen_etal_2015a,King_etal_2015}. 

\proj\ will transform the RM field by measuring broad-line time lags and RM-based SMBH masses for $\gtrsim 50$ AGNs at $1<z<4$ in the Deep tier, along with their host galaxy measurements from NIRCam imaging to explore co-evolution between SMBHs and galaxies \citep[e.g.,][]{Li_etal_2023}. By targeting F444W$\lesssim 23.5$ broad-line AGNs, \proj\ probes bolometric luminosities ($10^{44-46}\,{\rm erg\,s^{-1}}$) and BH masses ($10^{7-9}\,M_\odot$) that are ten times lower than ground-based programs at fixed redshift (Fig.~\ref{fig:rmagn}). Importantly, the 0.6--5.3\,\micron\ PRISM spectra allow simultaneous lag measurements for multiple lines across rest-frame UV-NIR, exploring the stratification of the broad-line region gas. It will be the first, and arguably the only feasible, RM program for \hbeta, \halpha, and Pa$\alpha$/Pa$\beta$ at $z>1$ and for these mass scales. These hydrogen lines have been shown to be the most reliable lines for RM measurements \citep{Shen_etal_2023}, mitigating systematic issues associated with UV broad lines such as \MgII\ and \CIV. 

These RM measurements from \proj\ will significantly improve the BH mass recipes based on single-epoch spectra towards the low-mass/high-$z$ regime, a critical step to gauge the reliability of such SMBH mass estimates for $z>6$ AGNs. The PRISM spectra have high S/N for the intended RM sample, allowing broad-line width measurements down to $\sim$ hundreds of ${\rm km\,s^{-1}}$ despite the low spectral resolution \citep{Kokorev2023,Greene2024_LRD}. Decomposing the narrow and broad components with PRISM for, e.g., \hbeta, is unnecessary: the width of the reverberating broad line is measured from the RMS spectrum, where the constant narrow-line flux is automatically removed. 

{Figure~\ref{fig:rmagn}} forecasts the RM yields for the targeted broad-line AGN sample anticipated from the Deep area, based on extrapolations of the high-redshift AGN LFs \citep[e.g.,][]{ShenX_LF2020}. Following earlier work \citep{Shen_etal_2015a,King_etal_2015}, we perform realistic simulations of mock AGN light curves and RM detection that capture the stochastic AGN variability \citep{MacLeod_etal_2012} and observing parameters (cadence, baseline and S/N). \proj\ will achieve 2--3\% epoch flux precision, using simultaneous NIRCam photometry and narrow \OIII\ line flux (always covered in PRISM) as intrinsic calibrators \citep{van_Groningen_1992}. This much higher precision, compared with the 5--10\% level achieved in ground-based multi-object spectroscopy \citep{Hoormann_etal_2019,Shen_etal_2023}, enables lag measurements with fewer spectroscopic epochs. The homogeneous spectroscopic sampling \citep[i.e., no seasonal gaps to cause aliasing lags, ][]{Shen_etal_2023} is another major improvement over ground-based programs. Despite the relatively small number of spectroscopic epochs, our tailored simulations (Fig.~\ref{fig:rmagn}) suggest a high success rate for lag measurements for the targeted RM AGNs. 

The targeted AGN sample has $i\lesssim 23.5-24$, bright enough for ground-based photometric monitoring to trace the most variable rest-UV continuum. Such ground-based monitoring with $\sim$ daily to weekly cadences over $\sim 9$ months/yr has been largely secured via external collaborations (\S\ref{sec:synergy}). These ground-based light curves not only provide much denser AGN light curves to facilitate RM lag measurements, but also supporting data for the time-domain sector of \proj. 

\subsection{Data Challenges and Unexpected Discoveries}\label{sec:ml}

\begin{figure}[!t]
\centering
\includegraphics[height=0.4\textwidth]{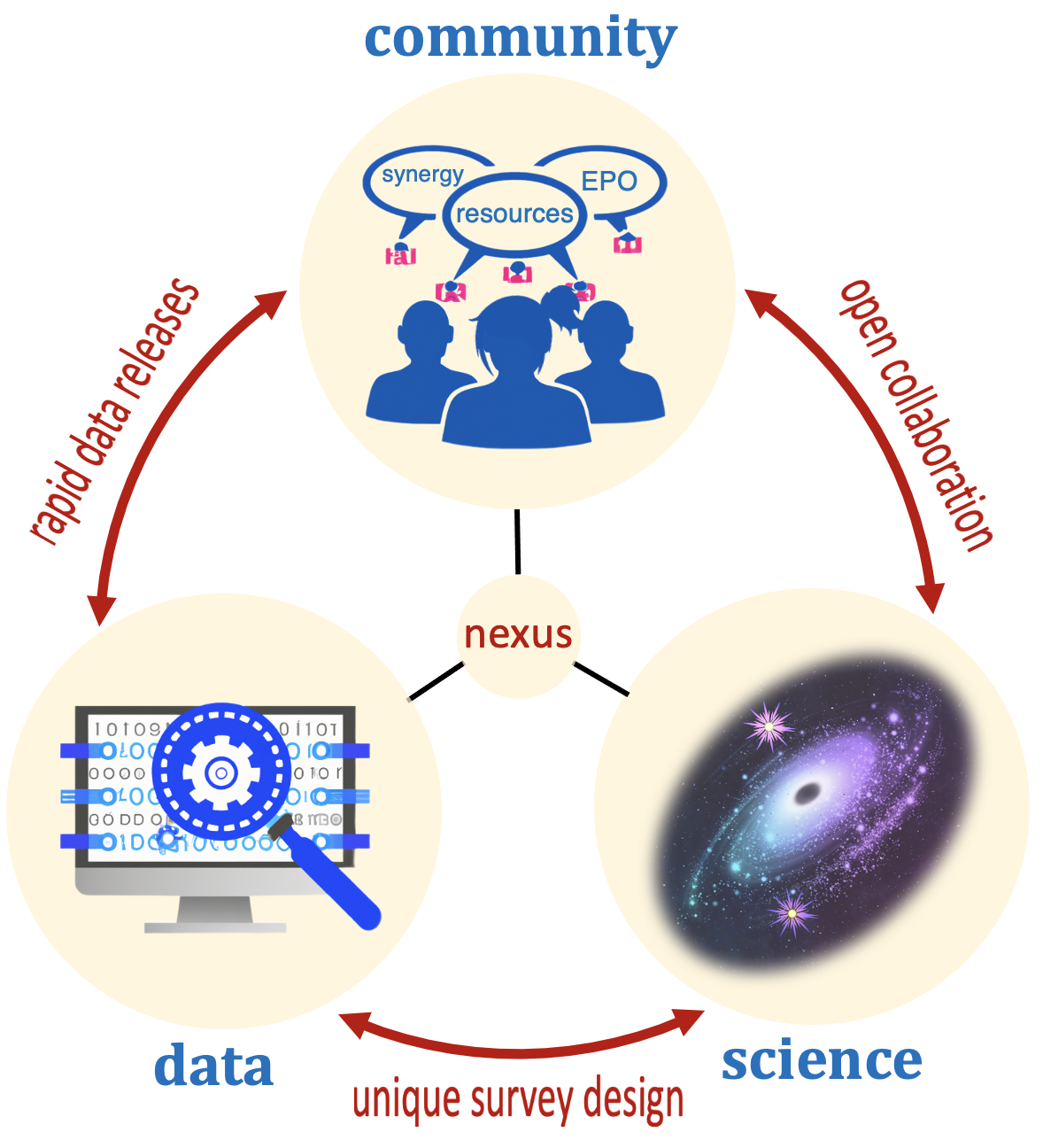}
\caption{The \proj\ ecosystem. The unique survey design facilitates the fusion between data and science for cutting-edge science topics and innovations in data techniques. Rapid data releases aid community engagement and follow-up work. The open collaboration model goes beyond the open-data policy for treasury programs and promotes genuine community-driven science advancements. }
\label{fig:eco}
%\vspace{-0.5cm}
\end{figure}

The comprehensive and multi-faceted data structure from three JWST instruments may produce unexpected discoveries (e.g., Solar System objects with highly-inclined orbits), and promote innovative ideas in data processing. Maximum extraction of science from these data poses data challenges that will prompt the community to think deeper in preparation for future missions. 

For example, there is significant room to improve on the extraction of WFSS grism spectra. One interesting problem is to extract WFSS spectra for bright transients that are contaminated or overwhelmed by neighbouring targets. Using multi-epoch WFSS observations at different PAs, it is possible to subtract such contamination from persistent sources (the transient in question will only appear in one epoch of WFSS). The three WFSS PAs will also provide spatial information for extended sources, and it may be possible to reconstruct spatially-resolved kinematic models for extended galaxies. Finally, with the deep accompanying multi-band photometry for source location, we will experiment WFSS extraction algorithms that model overlapping targets simultaneously. These improvements will prove valuable for all grism spectroscopic data from space-based facilities.

\subsection{Synergy and Community Engagement}\label{sec:community}

The \proj\ field, being in the continuous viewing zone and the Euclid Deep Field North (and the Euclid Ultra-deep Field therein), is likely to be covered by the primary High Latitude Time Domain field for Roman. The total area from this program, while much smaller than the Roman field-of-view, provides a nearly complete galaxy sample with unparalleled JWST depth and wavelength coverage. \proj\ will demonstrate the synergy between deep extragalactic science and time-domain science with cadenced imaging and spectroscopy, and the resulting much improved observing efficiency to address multiple key science goals. For example, a recent white paper from the HST/JWST Long-Term Monitoring Working Group \citep{Jha2024} emphasized the importance of monitoring the Ecliptic Pole regions, where the \proj\ field lies. Continued community investment of resources in the NEP region (specifically the \proj\ field) will significantly boost the legacy value of these JWST data. 

The \proj\ field is already densely sampled by the eROSITA X-ray telescope and the WISE Mid-IR telescope, since the NEP is also the continuous viewing zone for these satellites at the Sun-Earth L2. Although the depth of JWST is unmatched by eROSITA and WISE, X-ray and MIR light curves from these telescopes still provide useful information for the bright sources in \proj. There are also several ground-based programs that target the \proj\ field. For example, the HEROES survey \citep{heroes} has obtained deep Subaru/Hyper Suprime-Cam imaging for the NEP field, providing complete coverage of \proj. The Wide Field Survey Telescope \citep[WFST;][]{WFST} and the Young Supernova Experiment \citep[YSE;][]{YSE} are both photometrically monitoring the NEP field that fully covers \proj. In particular, the 2.5~m WFST facility is at a favorable geo-latitude with $\sim 9$~months of annual visibility for the NEP. WFST is monitoring the NEP field with nearly daily cadence, and these ground-based light curves will complement the \proj\ observations.  

\proj\ will foster an ecosystem for community-shared data, knowledge transfer to other programs, and innovations in data processing techniques (Fig.~\ref{fig:eco}). With the largest spectroscopic sample from JWST, it promises significant improvements in spectroscopic training samples for photo-z estimation. Unlike other heavily invested extragalactic fields (e.g., GOODS-N, GOODS-S, COSMOS), the \proj\ field is a new JWST field, adding $\sim 0.1\,{\rm deg^2}$ towards a total collective {$\sim 1\,{\rm deg^2}$} legacy area. The added sky area and statistics further mitigate cosmic variance. The improved photo-z algorithms will be applied to other extragalactic fields, where deep spectroscopy is still relatively sparse, to significantly amplify the legacy value of JWST imaging data.

Finally, \proj\ envisions an open-collaboration model (see detailed scopes in \S\ref{sec:open}), and invites the community to share resources. The team does not claim ownership of data or science, nor does it require proprietary access to other data sets such as Euclid. \textbf{Rather, the data and science are for everyone}. The team is committed to delivering data products in a timely fashion (\S\ref{sec:roadmap}) to enable rapid community investigations and investments of follow-up resources. 

\begin{table*}[]
\caption{Summary of Depths}\label{tab:depth}
\centering
%\scalebox{0.9}{
\begin{tabular}{lcccccccccccc}
\hline\hline
 &  & Spec & Spec & F090W & F115W & F150W & F200W & F356W & F444W & MIRI & MIRI  & MIRI  \\
&  & F322W2 & F444W & (SW) & (SW) & (SW) & (SW) & (LW) & (LW) & F770W & F1000W & F1280W  \\
\hline
{\bf Wide} & Epoch1  & 1.2e-17 & 7.0e-18 & 27.4 & 27.3 & 27.2 & 28.3 & 27.7 & 27.9 & 25.0 & 24.6 & 24.0  \\
(WFSS) &   & (21.7) & (21.6) &  &  & &  &  &  &  &  &   \\
2.4-5\micron & Final & 6.0e-18 & 3.3e-18 & 27.8 & 27.9 & 27.9 & 29.0 & 28.4 & 28.3 & - & - & - \\
 &   & (22.3) & (22.2) &  & & &  &  &  &  &  &   \\
\hline
 &  & 3.2 \micron & 4.4 \micron & - & - & F210M & F200W & F360M & F444W & - & F1000W & -  \\
 \hline
{\bf Deep} & Epoch1 & 1.2e-18  & 1.1e-18 & - & - & 27.7 & 27.7 & 28.0 & 27.7 & - & 24.3 & - \\
(PRISM) &  & (26.6) & (25.6) &  &  &  & & &  & & & \\
0.6-5.3\micron & Final & 2.4e-19 & 2.3e-19 & - & - & 28.3 & 29.4 & 28.6 & 29.3 & - & 24.9 & - \\
  &  & (28.3) & (27.3) &  &  &  & & &  & & \\
\hline
\hline
\end{tabular}
%\tablecomments{Depth table}
%}
\tablecomments{\raggedright Emission-line flux limits (from spectroscopy; cgs units) and photometric depths (from imaging) are 5$\sigma$ for point sources. The magnitude limit in parentheses corresponds to the continuum-dominated target brightness to reach average spectral S/N$\sim 3$~pixel$^{-1}$ across the bandpass (F322W2 and F444W) -- this threshold is necessary to measure a redshift per experience from spectroscopic surveys such as the SDSS. {The depths of NIRCam WFSS are estimated based on real observations in FRESCO and CONGRESS surveys.} }
\end{table*} 

\begin{table*}[]
\caption{Source counts (targeted) }\label{tab:counts}
\centering
%\scalebox{0.8}{
\begin{tabular}{lccccccc}
\hline\hline
 & $1<z<6$ gal & $z>6$ gal & $z>9$ gal & {$z>5$ line emitter} & {RM AGN} & $z>3$ SN Ia & Cold stars  \\
\hline
{\bf Wide imaging} & 1.2e5 & 4000 & 80 & - & - & 30 & 300 \\
{\bf WFSS 2.4-5\,\micron} & 2500--{3800} & - & - & 1000 & - & -  & 15 \\
{\bf PRISM 0.6-5.3\,\micron} & $>5000$ & 100 & - & - & {50--100} & 20 & 40 \\
\hline
\hline
\end{tabular}
%}
%\raggedright 
\tablecomments{Galaxy counts based on simulated samples \citep{JAGULAR_mock} and verified with actual counts in COSMOS-Web (Fig.~\ref{fig:depth} right). RM AGN counts based on the predicted range from different LF extrapolations \citep{ShenX_LF2020,Ueda2014_obscured}. SN Ia counts based on expected rates (Fig.~\ref{fig:sn}). Brown dwarf counts based on population synthesis and evolutionary models \citep{2023arXiv230812107B}. For $1<z<6$ galaxies with spec-z from WFSS, the lower bound is from the expected galaxy counts at F444W$<22$ (Fig.~\ref{fig:counts}), and the upper bound is scaled from the observed numbers in CONGRESS \& FRESCO (X. Lin \& F. Sun, private communications).}
\end{table*}

\begin{table}[]
\caption{Exposure Times}\label{tab:expt}
\centering
%\scalebox{0.9}{
\begin{tabular}{cc}
\hline\hline
Filter/Grating & Exposure time (sec) \\
\hline
\multicolumn{2}{c}{\textbf{Wide NIRCam Imaging}}\\
\hline
F090W & 934.1 (1868.2)\\
\hline
F115W & 622.7 (1868.2)\\
\hline
F150W & 311.4 (1245.5)\\
\hline
F200W & 1245.5 (3736.4)\\
\hline
F356W & 311.4 (1245.5)\\
\hline
F444W & 934.1 (1868.2)\\
\hline
\multicolumn{2}{c}{\textbf{Wide NIRCam WFSS}}\\
\hline
F322W2 & 622.7 (1868.2)\\
\hline
F444W & 1245.5 (3736.4)\\
\hline
\multicolumn{2}{c}{\textbf{Wide MIRI Imaging}}\\
\hline
F770W & 957.4/1259.9\\
\hline
F1000W & 1259.9/957.4\\
\hline
F1280W & 957.4/638.3\\
\hline
\multicolumn{2}{c}{\textbf{Deep NIRCam Imaging}}\\
\hline
F200W & 622.7 (11209.2)\\
\hline
F210M & 1245.5 (3736.4)\\
\hline
F360M & 1245.5 (3736.4)\\
\hline
F444W & 622.7 (11209.2)\\
\hline
\multicolumn{2}{c}{\textbf{Deep MIRI Imaging}}\\
\hline
F1000W & 638.3 (1914.8)\\
\hline
\multicolumn{2}{c}{\textbf{Deep NIRSpec MOS}}\\
\hline
PRISM & 1283.8 (23108.8)\\
\hline
\end{tabular}
\tablecomments{Exposure times of imaging and spectroscopy of NEXUS-Wide and NEXUS-Deep for single epochs and for final coadds (shown in parentheses). For Wide MIRI parallel imaging, the filter combination alternates among the three epochs, and each filter has two different exposure times. }
\end{table}

\begin{figure*}[!t]
\centering
\includegraphics[width=0.48\textwidth]{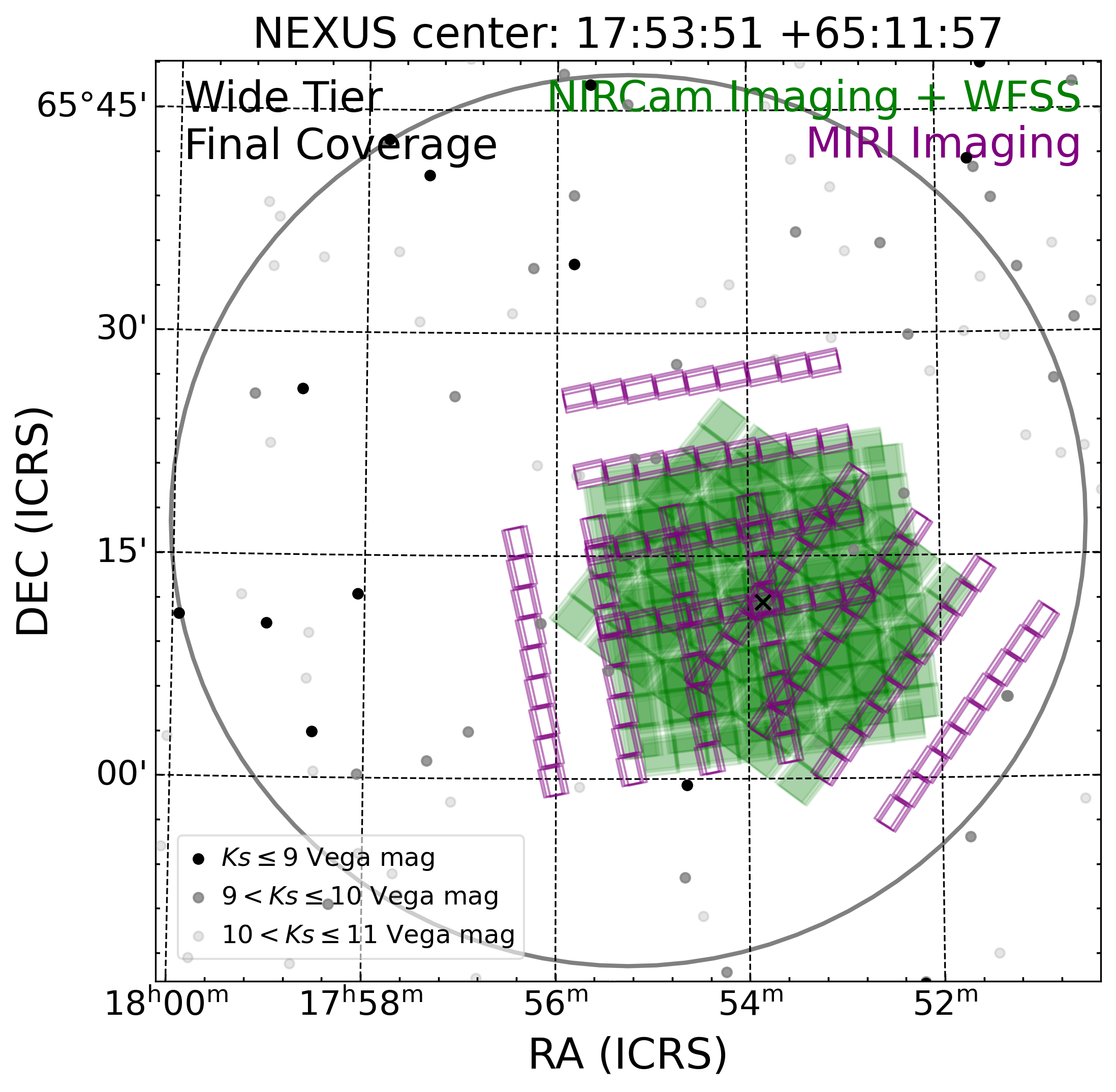}
\includegraphics[width=0.48\textwidth]{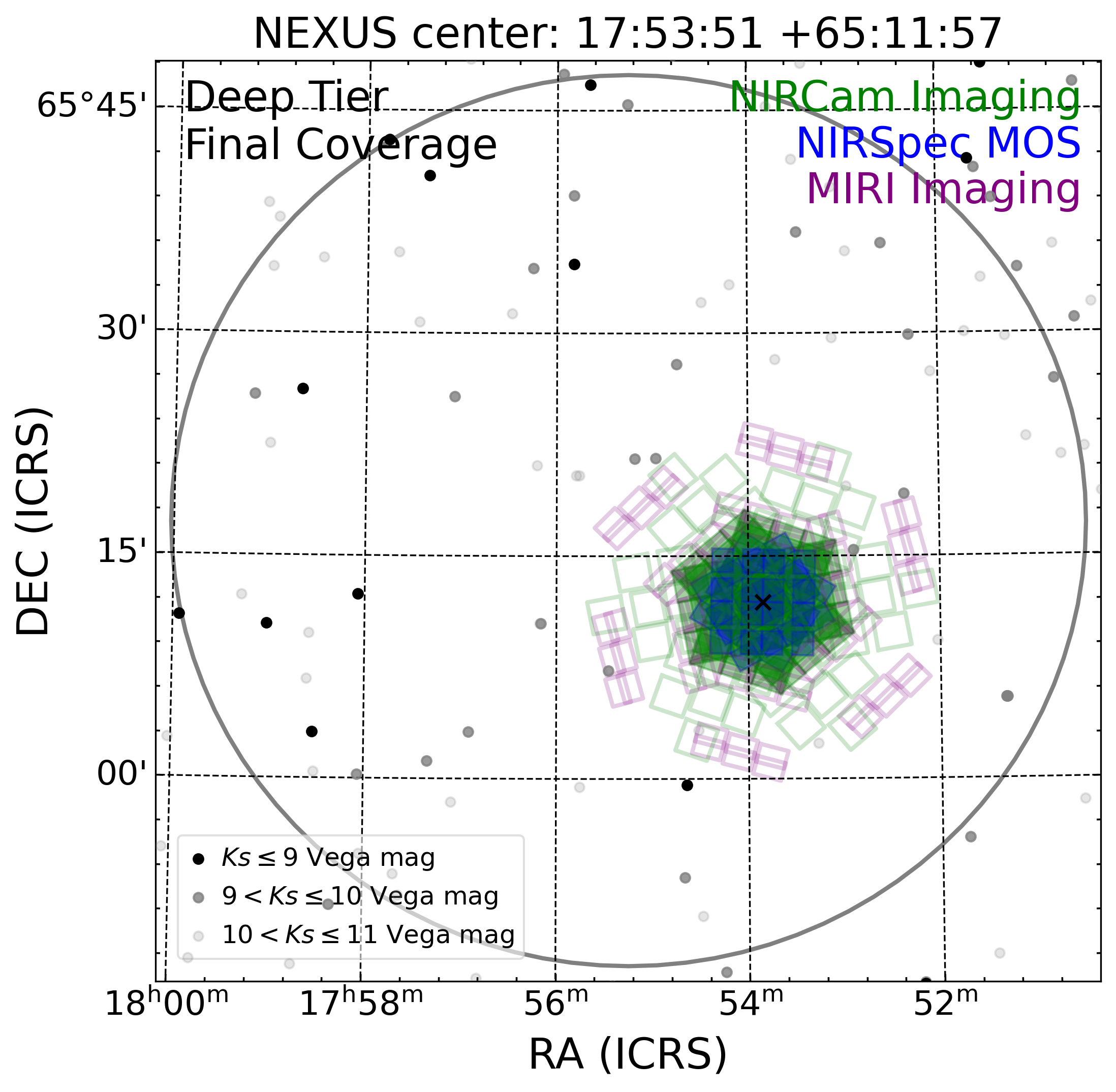}
\caption{Survey layouts with all epochs combined. For each labeled instrument (e.g., NIRCam, MIRI), the footprint is identical for all filters. The center of the plot is the central pointing of the Euclid Self-calibration field (shown in the black circle), which itself is within the Euclid Deep Field North \citep{euclid}. The Wide tier (left) is centered in a local stellar density minimum, with a primary coverage (NIRCam WFSS+Imaging) of $\sim 400\,{\rm arcmin}^2$. The Deep tier has a primary coverage (NIRSpec MOS+NIRCam imaging) of $\sim 50\,{\rm arcmin}^2$, with additional $\sim 230\,{\rm arcmin}^2$ parallel NIRCam imaging in two medium bands (F210M+F360M). The medium band coverage is fully within Wide. In both panels, filled blocks are primary instrument coverage and open blocks are parallel instrument coverage. Bright stars are marked as colored dots. }
\label{fig:footprint}
\end{figure*}

\section{Program Implementation}\label{sec:design}

\noindent\fire{Overview of survey design}\quad The \proj\ field is near the NEP and lies completely within the Euclid Deep Field North \citep[EDFN;][]{euclid}. In fact, the \proj\ field is fully covered in the Euclid Self-calibration field, which will eventually become the Euclid Ultra-Deep Field with $\sim$monthly visits \citep{euclid}. The synergy between Euclid and \proj\ is excellent, both in terms of complementary wavelength coverage (Fig.~\ref{fig:depth}) and temporal sampling.  

We had carefully considered a previous JWST Time-Domain Field \citep[TDF;][]{TDF_2018}, which unfortunately lies outside EDFN and has limited area. There is essentially no difference between the two fields in stellar density and extinction. While the Euclid data are not mandatory on field selection, the added uniform 0.9--2~\micron\ spectroscopic coverage and $\sim$monthly repeated photometry from Euclid for the entire Wide tier is highly desirable. There are also ongoing ground-based, intensive monitoring programs covering EDFN and thus \proj\ \citep[e.g.,][]{YSE, WFST}  but not the TDF, which are more crucial to our time-domain component. An even stronger argument to select a new NEP field is to provide expanded area coverage to further mitigate cosmic variance, and the spectroscopic training information is transferable to the other imaging fields as well. 

The Wide tier aims to reach a final imaging depth of $\sim 28-29$ in most NIRCam SW/LW broad-band filters and nearly complete WFSS grism coverage over a contiguous area of $\sim 0.1\,{\rm deg^2}$ (\textbf{primary coverage 394~arcmin$^2$}). These goals can be achieved with a $4\times9$ mosaic NIRCam tiling, requiring total charged time of $\sim 180$~hrs. Thus the Wide tier observations are divided in three epochs across three cycles (with each epoch covering the full area). The three Wide epochs will provide a $\Delta t\sim 1.5$~yr, and three nominal PAs (PA1, PA1$+135^{\circ}$, PA1$+270^{\circ}$) for the WFSS wavelength dispersion direction. For each epoch+pointing, simultaneous NIRCam SW imaging is obtained with WFSS grism, with additional direct SW+LW (F356W+F444W) imaging for WFSS source tracing. MIRI imaging is taken as parallel. We choose F322W2+F444W filters for WFSS: F444W is most critical to high-$z$ line-emitter identification; F322W2 is an upgraded filter to replace F356W with extended wavelength coverage down to $2.4$~\micron. Targets overlap and confusion will be resolved with three dispersion directions and source detection in F356W+F444W direct imaging. In particular for the identification of line emitters, target overlap is not a concern. 

For the Deep tier spectroscopy, we switch to NIRSpec/MOS with PRISM for its much higher sensitivity and wider wavelength coverage (\textbf{primary coverage 49~arcmin$^2$}). For each epoch, we use 4 NIRSpec pointings with parallel NIRCam imaging. Because the NIRCam parallel has different FoV than NIRSpec, we require separate primary F200W+F444W NIRCam imaging at each pointing to cover the PRISM area, paired with MIRI parallel imaging. These time-resolved NIRCam imaging observations are optimal for high-$z$ SNe searches, and reference NIRCam images are already available from Wide. The Deep tier will be re-visited on a $\sim$2-month cadence, providing 6 different PAs each cycle to maximize the flexibility of MOS target allocation. Initial slew overhead for each epoch is negligible compared with the total time charge, and the cadenced observations open the window for systematic time-domain explorations. The single-epoch imaging and spectroscopic depths are sufficient for key time-domain science, and provide the ability to observe nearly all targets at F444W$<25.7$ with PRISM. Each epoch (4 NIRSpec pointing) can accommodate more than $\sim 650$ MOS targets, among which $\sim 250$ are reserved for fixed targets (monitoring targets and high-priority targets requiring deep spectroscopy, e.g., $z>9$ candidates). The remaining $\sim 400$ slots are used for different targets each epoch, providing $\sim 7200$ MOS slots to cover most of the anticipated $<10^4$ distinct targets at F444W$<25.7$ within Deep (Fig.~\ref{fig:counts}). The exact numbers of MSA allocations are to be determined, and it is possible we will achieve an even higher spectroscopic completeness for the flux-limited sample in Deep.  

Accompanying the NIRSpec observations, parallel MIRI and NIRCam imaging adds more filters to maximize the science, e.g., to improve object classification and photometric color measurements. In particular, the $\sim 230\,{\rm arcmin}^2$ parallel NIRCam imaging (in Deep) in two medium bands (F210M+F360M) are tuned for ultra-faint line-emitter searches at specific redshifts. For example, this filter combination can effectively select objects at $z=4.6$ with both \OII~3727 and H$\alpha$, \OIII\ emitters at $z\sim 6.2$, H$\beta$ emitters at $z\sim 6.4$, and \OII\ emitters at $z\sim 8.7$. Finally, these parallel observations are cadenced, providing valuable variability information at these additional wavelengths. 

The detailed survey layouts are shown in Fig.~\ref{fig:footprint}. The wavelength coverage and point-source sensitivity are shown in Fig.~\ref{fig:depth}, and the depths are summarized in Table~\ref{tab:depth}. Expected source counts are compiled in Table~\ref{tab:counts}. We have optimized instrument/filter combinations and exposure times to maximize science utility and to minimize observational overhead. The total allocated time for the Wide and Deep tiers are 183~hrs and 183.6~hr, respectively. Each Wide epoch will cost $\sim 65$~hrs and each Deep epoch will cost $\sim 11$~hrs. The individual exposure times for each observing mode and each filter are listed in Table~\ref{tab:expt}.

\begin{figure}[]
\centering
\includegraphics[width=0.445\textwidth]{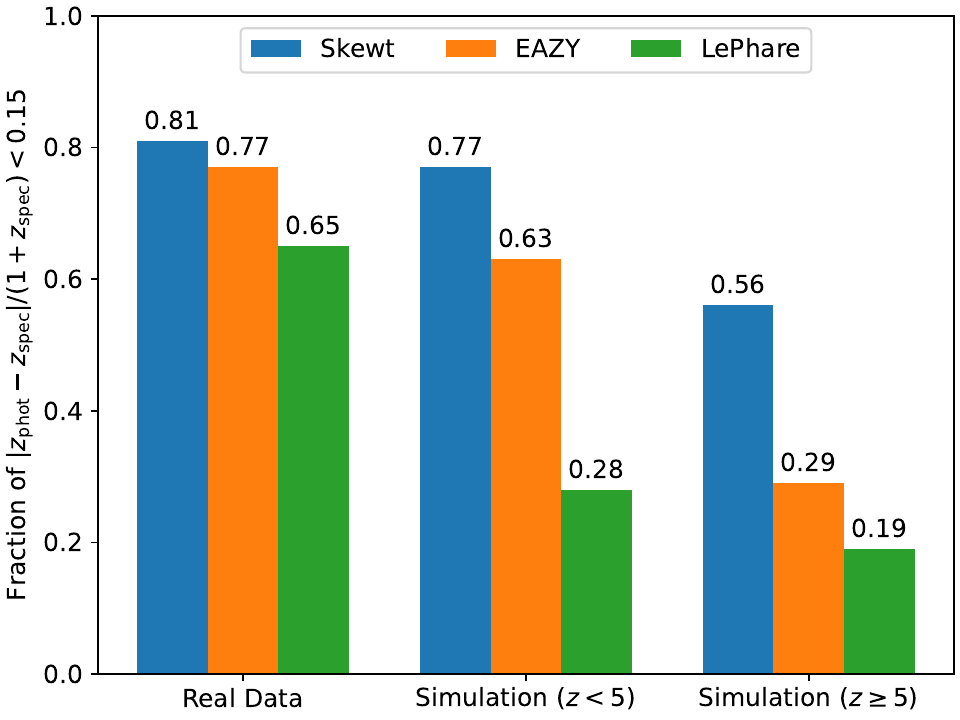}
\caption{Performance of galaxy photo-z selection using Skew-t \citep{skewt}, EAZY \citep{eazy} and LePhare \citep{lephare} with either real data (from the spectroscopic sample in COSMOS-Web; dominated by $z<5$ galaxies) or simulated galaxies that cover a much broader redshift range \citep{JAGULAR_mock}. Photometric bands are limited to the broad-band filters in \proj\ and with the single-epoch depths as indicated in Fig.~\ref{fig:depth} and Table~\ref{tab:depth}. The optimal photo-z selection will be applied to the first Wide epoch to perform target selection for NIRSpec/MOS.  }
\label{fig:photoz}
%\vspace{-0.5cm}
\end{figure}

\noindent\fire{Scheduling of Wide and Deep epochs}\quad Due to a Cycle-3 scheduling requirement to maximize observing time in the May 20--Aug 30, 2025 window, the first Wide epoch is split into two parts. The first part, covering the central {$2\times 5$} NIRCam/WFSS pointings as well as the full Deep area, will be observed in Sep 2024. The remaining Wide Epoch 1 will be observed during May--Jun 2025. The Deep epochs will start right after the completion of the first Wide epoch near the end of Cycle 3. This slight modification of scheduling compared with our original plan has the following benefits, while complying with the Cycle 3 scheduling requirements: (1) the early NIRCam/WFSS observation of the central Wide (and full Deep) area provides reference images and source catalogs for NIRSpec/MSA target selection in Deep, with sufficient lead time; (2) the effective baseline of \proj\ is extended by nearly a year, facilitating variability studies and the planning for coordinated programs; (3) the early observations of a small fraction of Wide area will test our program design, especially the performance of the F322W2 WFSS filter, which is infrequently used in other extragalactic NIRCam/WFSS programs. Although observing part of the Wide Epoch 1 at an earlier time results in a different PA, it is not a major issue as we can adjust future Wide epochs to avoid completely identical PA for the central area. 

Overall, the three Wide epochs will be separated by a long $\Delta t\approx 1.5$~yr, and also stretch the full duration of Cycles 3--5. On the other hand, the Deep epochs will be executed on a $\sim$2-month cadence ($\pm 2$~weeks is possible due to scheduling) since May 2025, and will extend into Cycle 6. While we will have the necessary photometry for the bulk of NIRSpec target selection immediately following Wide Epoch 1, we expect regular tweaks to our MSA targets for successive Deep epochs, e.g., to replace failed broad-line AGN targets, to add new emerging transient targets, and to utilize imaging data from additional Wide+Deep epochs.

We will deploy state-of-the-art photometric selection for the NIRSpec MSA targets. We have compared different photo-z selection algorithms \citep{skewt,eazy,lephare} using the expected broad-band photometry (Table~\ref{fig:depth} and Fig.~\ref{fig:depth}), benchmarked upon both real (from the spectroscopic sample in COSMOS-Web) and simulated data \citep{JAGULAR_mock}. As shown in Fig.~\ref{fig:photoz}, the best-performing (so far) photometric selection produces a high completeness ($56\%$) for $z>5$ galaxies, with a comfortably low ($<15\%$) contamination rate from lower-$z$ interlopers, meaning we will have a high success rate for $z>5$ targets. Similarly, our photo-z selection recovers $\sim 90\%$ of the intended RM broad-line AGNs with an acceptable ({$<30\%$}) over-targeting rate -- these failed RM AGN targets will be removed from successive epochs. We are still improving our photometric selection and photo-z estimation in deep JWST fields (Yang, Q.~et~al., in prep), and expect to improve target selection as the project progresses. In any case, we have more than enough MSA slots to accommodate the vast majority of a flux-limited sample, and the prioritized target selection is mostly for fainter $z\gtrsim 6$ galaxy targets or transients. 

\begin{figure*}[]
\centering
\includegraphics[width=0.98\textwidth]{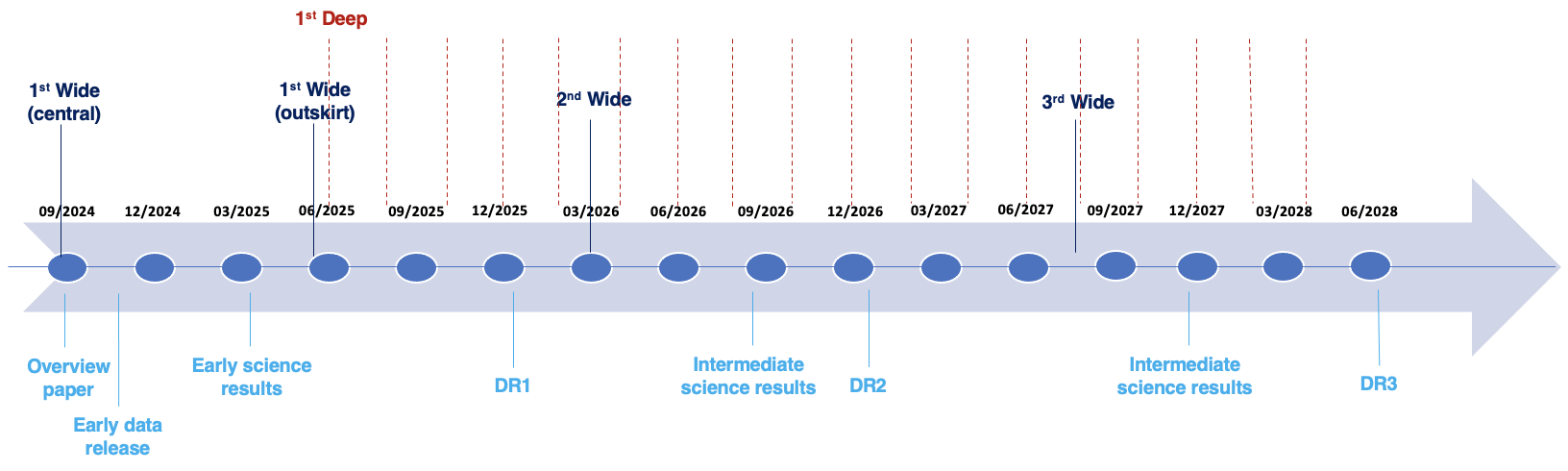}
\caption{Preliminary timeline and milestones of \proj\ through June 2028. The first observations (Wide Epoch 1 central) will be executed during Aug/Sep 2024, and the bulk of the program will start in May 2025. Official Data Releases 1 and 2 are scheduled 6 months after the completion of the prior cycle, while DR3 is scheduled in mid 2028 after all Deep epochs are taken. Rapid quick-reduction releases are anticipated within 4 weeks after the observations are taken. Science results will be produced as the project progresses, and we anticipate bulk science production at several milestones along the project timeline. }
\label{fig:timeline}
%\vspace{-0.5cm}
\end{figure*}

\section{Data Production and Collaboration}\label{sec:synergy}

\subsection{Data Production Roadmap}\label{sec:roadmap}

To maximize the efficiency of planned observations and to enable rapid community follow-up, the team will immediately produce and publicly release reduced data following each epoch within $\sim 4$ weeks. Six months after the completion of each cycle, the team will publicly release fully calibrated data for that cycle, in the form of annual data releases. Value-added products, such as mosaic images, source catalogs, classifications, merged multi-wavelength photometry, and web tools, etc., will be interlaced between data releases. Given the scheduling requirements of the program, the team will work closely with STScI and JWST instrument teams to ensure smooth implementation and data reduction. {Fig.~\ref{fig:timeline} shows the current project timeline and milestones.} The latest updates on data production schedules can be found on the \proj\ website. 

\subsection{Open Collaboration Model}\label{sec:open}

The raw and low-level auto-processed JWST data from \proj\ are immediately public given its treasury program nature. The \proj\ team strongly encourages the community to use these data for immediate investigations. To facilitate community usage of \proj\ data, the team is committed to rapid releases of quick-reduction data following each epoch (\S\ref{sec:roadmap}). Improved calibrations will be released in annual data releases, along with value-added data products spread over the course of the survey. The \proj\ team also encourages coordinated community follow-up programs, and will provide technical information and coordination. 

There is significant synergy between \proj\ and other large survey programs (\S\ref{sec:community}). The \proj\ team will seek potential collaborations with these external teams to maximize the science in the \proj\ field. The detailed scope of these collaborations will be formulated in the form of memoranda of understanding. 

Given the tremendous amount of \proj\ data and the diverse range of science topics these data will address, the core team will build the \proj\ expanded team to work on the data. This open-collaboration model goes beyond the open-data policy for treasury programs, in that expanded \proj\ team members can have access to a much broader suite of supporting data assembled and curated by the team. In addition, expanded team members can propose important NIRSpec/MSA targets, lead science papers, and collaborate on additional follow-up observing programs. 

While the number of expanded team members is limited by management capacities, we do expect to expand the core team significantly over the course of the project. In general, we expect expanded team members bring unique expertise and resources to the shared pool of \proj. Examples include, but are not limited to, major observing resources that complement the key science of \proj, which could be large-aperture ground facilities, dense ground-based optical-IR light curves, or triggers of high-profile transients for NIRSpec MOS; advanced expertise on JWST data reduction, data analysis tools, and theoretical perspectives. The \proj\ Executive Committee will review individual cases and approve such requests on an annual basis (first call likely in Spring 2025). Team members must abide by the Principles of Operations of the project to follow professional conduct, which is being formulated at the time of writing. 

\section{Summary}\label{sec:sum}

JWST has opened a new era in exploring the faint and distant Universe. In many ways, JWST observations are transforming our knowledge about the history of the cosmos and everything within. The rich array of observing modes and unprecedented instrument capabilities of JWST have enabled exciting programs that are revealing the mysteries of the Universe. 

Riding on the success of previous large JWST programs, the Multi-Cycle (Cycles 3--5) JWST Treasury program \proj\ will perform one of the most ambitious extragalactic surveys with JWST. Massive and uniform spectroscopy, and systematic time-domain investigations, are the first two pillars of \proj. Multi-field synergy and community science add to the third pillar. When combined, \proj\ will enable an exceptionally broad spectrum of science topics, from the earliest galaxies and SMBHs, to the farthest transients in the distant Universe. 

This overview paper introduces the \proj\ project: science motivation and program design. We also provide a timeline for the project, and additional information for community engagement. In short, \proj\ will acquire deep NIRCam+MIRI photometry and a large number of NIRCam/WFSS and NIRSpec/MSA spectra for a two-tier area coverage. The Wide tier ($\sim 400\,{\rm arcmin^2}$) performs primary NIRCam/WFSS spectroscopy for three epochs distributed over three years. The Deep tier ($\sim 50\,{\rm arcmin^2}$) is within Wide, and performs primary NIRSpec/MSA multi-object spectroscopy for $\sim 10^4$ targets, over 18 epochs distributed on a 2-month cadence. The \proj\ field is around the North Ecliptic Pole, with unlimited JWST visibility, and is within the Euclid Deep Field North to maximize synergy. 

We describe the relevance of \proj\ in the landscape of JWST studies for galaxy evolution and time-domain science, with highlighted key science areas. With the open-collaboration model and community shared resources, we hope that the \proj\ field will become a new benchmark field for extragalactic science in the coming decade. Together, we look forward to the exciting discoveries!   

\acknowledgments

We thank the STScI JWST scheduling group and the Program Coordinator, Alison Vick, for helping with the scheduling of the program. Based on observations with the NASA/ESA/CSA James Webb Space Telescope obtained from the Barbara A. Mikulski Archive at the Space Telescope Science Institute, which is operated by the Association of Universities for Research in Astronomy, Incorporated, under NASA contract NAS5-03127. Support for Program numbers JWST-GO-02057, JWST-AR-03038, and JWST-GO-05105 (YS, MZ and JL) was provided through a grant from the STScI under NASA contract NAS5-03127. GN gratefully acknowledges NSF CAREER grant AST-2239364, supported in-part by funding from Charles Simonyi, support from NSF AST-2206195, and NSF OAC-2311355, DOE support through the Department of Physics at the University of Illinois, Urbana-Champaign (13771275), and support from the HST Guest Observer Program through HST-GO-16764 and HST-GO-17128 (PI: R. Foley).

%\clearpage

\bibliographystyle{aasjournal}
\bibliography{refs.bib}

\end{document}